\documentclass[12pt,a4paper]{article}

\usepackage{textcomp}
\usepackage[latin1]{inputenc}
\usepackage{pstricks,pst-node,pst-text}
\usepackage{amsmath}
\usepackage{graphicx}

\newcommand{\be}{\begin{equation}}
\newcommand{\ee}{\end{equation}}
\newcommand{\bea}{\begin{eqnarray}}
\newcommand{\eea}{\end{eqnarray}}
\newcommand{\beqn}{\begin{eqnarray}}
\newcommand{\eeqn}{\end{eqnarray}}
\newcommand{\ba}{\begin{array}}
\newcommand{\ea}{\end{array}}

\newcommand{\ben}{\begin{enumerate}}
\newcommand{\een}{\end{enumerate}}
\newcommand{\nn}{\nonumber}
\newcommand{\crn}{\nonumber \\}

\newcommand{\noi}{\noindent}

\newcommand{\la}{\lambda}

\newcommand{\ga}{\gamma}

\newcommand{\eps}{\varepsilon}

\newcommand{\fr}{\frac}
\newcommand{\bc}{\begin{center}}
\newcommand{\ec}{\end{center}}

\newcommand{\ep}{\epsilon}

\newcommand{\ra}{\rightarrow}

\newcommand{\beq}{\begin{equation}}
\newcommand{\eeq}{\end{equation}}
\def\slashc{c\kern -.400em {/}}
\def\slashp{p\kern -.400em {/}}
\def\slashq{q\kern -.450em {/}}
\def\slashL{L\kern -.450em {/}}
\def\slashcl{\cl\kern -.600em {/}}
\def\slashr{r\kern -.450em {/}}
\def\slashk{k\kern -.500em {/}}
\def\slashep{\epsilon\kern -.450em {/}}
\def\slashpbar{\bar{p}\kern -.450em {/}}

\newcommand{\alio}{{\cal A}(\hat\la)}
\newcommand{\ali}{{\cal A}(\la_1,\la_2;\la_3,\la_4)}
\newcommand{\aliq}{{\cal A}(\la_1,\la_2;\la_3,\la_4;q_1,q_2)}
\newcommand{\aliqp}{{\cal A}(\la_1,\la_2;\la_3,\la_4;q^\prime_1,q^\prime_2)}
\def\slashepi{\epsilon_i\kern -.720em {/}}
\def\slashpi{p_i\kern -.600em {/}}

\begin{document}

\begin{titlepage}

\vspace*{0.1cm}\rightline{LAPTH-1216}
\vspace*{0.1cm}\rightline{CERN-PH-TH/2007-221}

\vspace{1mm}
\begin{center}

{\Large{\bf Leading Yukawa corrections to Higgs production
associated with a tagged bottom anti-bottom pair in the Standard
Model at the LHC}}

\vspace{.5cm}

F.~Boudjema${}^{1)}$ and  LE Duc Ninh${}^{1,2)}$

\vspace{4mm}

{\it 1) LAPTH, Universit\'e de Savoie, CNRS,  \\
BP 110, F-74941
Annecy-le-Vieux Cedex, France}
\\ {\it
2) CERN, Theory Division, \\
CH-1211 Geneva 23, Switzerland}
\\

\vspace{10mm} \abstract{Considering the large value of the top
Yukawa coupling, we investigate the  leading one-loop Yukawa
electroweak corrections that can be induced by the top quark in a
process such as Higgs production in association with a tagged
bottom anti-bottom pair at the LHC. At NLO these contributions are
found to be small at the LHC both for the total cross section and
for the distributions. In the limit of vanishing bottom Yukawa
coupling where the LO contribution vanishes, the process can still
be induced at one-loop through the top quark transition. Though
this contribution which can be counted as part of the NNLO
correction is small for Higgs masses around $120$GeV, it quickly
picks up for higher Higgs masses. This contribution represents the
rescattering of the top quarks and their decay into $W$'s leading
to Higgs production through $WW$ fusion.}

\end{center}

\normalsize
\end{titlepage}

\section{Introduction}
The most important goal of the LHC is the discovery of the Higgs
and the concomitant study of the mechanism of electroweak symmetry
breaking. Especially if no new phenomenon is unravelled through
the direct production of new particles, the study of the Higgs
properties such as its self-couplings and couplings to the other
particles of the standard model (SM) will be crucial in order to
establish the nature of the scalar component of the model. In this
respect most prominent couplings, in the SM, are the Higgs, the
top, and to a much lesser degree the bottom, Yukawa couplings. The
top Yukawa coupling is after all of the order of the strong QCD
coupling and plays a crucial role in a variety of Higgs related
issues. The dominant mechanism for Higgs production at the Large
Hadron Collider (LHC) is the gluon fusion process, which
incidentally is initiated through a top loop. Electroweak gauge
boson fusion and $W/Z H$ associated production \cite{djouadi_H1}
are also of importance. Higgs production associated with heavy
quarks like the top or bottom quark is not considered as a
discovery channel because of its small total cross section, the
top suffering further from a complicated final state topology.
However, if one wants to determine the bottom-Higgs Yukawa
coupling, $\la_{bbH}$, then Higgs production associated with a
bottom anti-bottom pair could provide a direct measurement of this
coupling. In the minimal supersymmetric standard model, MSSM, the
bottom   Yukawa coupling is enhanced by a factor $\tan\beta$, the
ratio of the vacuum expectation values of the two Higgs doublets.
For high $\tan\beta$ and not too large Higgs masses this provides
an important discovery channel for the supersymmetric Higgses.

The next-to-leading order (NLO) QCD correction to $pp\to
b\bar{b}H$ has been calculated by different groups relying on
different formalisms. In a nut-shell, in the five-flavour scheme
(5FNS)\cite{Barnet-Haber-Soper,Willenbrock_bbh_original}, use is
made of the bottom distribution function so that the process is
approximated (at leading order, LO) by the fusion $b \bar b \ra H$.
This gives an approximation to the inclusive cross section
dominated by the untagged low $p_T$ outgoing $b$ jets. If only one
final $b$ is tagged, the cross section is approximated by $g b \ra
bH$. The four flavour scheme (4FNS) has no $b$ parton initiated
process but is induced by gluon fusion $gg \ra b \bar b H$, with a
very small contribution from the light quark initiated process
$q\bar{q}\to b\bar{b}H$\footnote{In fact $q\bar{q}\to b\bar{b}H$
is dominated by $q\bar{q}\to HZ^*\to b\bar{b}H$ and does not
vanish for vanishing bottom Yukawa coupling. However this
contribution should be counted as $ZH$ production and can be
excluded by imposing an appropriate cut on the invariant mass of
the $b\bar{b}$ pair.}. Here again the largest contribution is due
to low $p_T$ outgoing $b$'s which can be accounted for by gluon
splitting into $b\bar b$. The latter needs to be resummed and
hence one recovers most of the 5FNS calculation while retaining
the full kinematics of the reaction. QCD NLO corrections have been
performed in both
schemes\cite{Willenbrock_bbh_original,dittmaier_bbH,dawson_bbH,dawson_bH_bbH}
and one has now reached a quite good agreement\cite{LH05-Higgs}.\\
The 5FNS approach, which at leading order is a two-to-one process
has allowed the computation of the NNLO QCD
correction\cite{WillenbrockNNLObbh,HarlanderKilgore} and very
recently the electroweak/SUSY (supersymmetry)
correction\cite{Dittmaier_bbH_susy} to $b \bar b \ra \phi$, $\phi$
any of the neutral Higgs boson in the MSSM. SUSY QCD corrections
have also been performed for $gg \ra b \bar b
h$\cite{Deltamb-bbH,Hollik_susy_QQh} where $h$ is the lightest
Higgs in the MSSM as well as to  $g b \ra b
\phi$\cite{Dawsonbbh_susy}. \\

\noindent In order to exploit this production mechanism to study
the Higgs couplings to $b$'s, one must identify the process and
therefore one needs to tag both $b$'s, requiring somewhat large
$p_T$ $b$. This reduces the cross section but gives much better
signal over background ratio. For large $p_T$ outgoing quarks one
needs to rely on the 4FNS to properly reproduce the hight $p_T$
$b$ quarks. The aim of this paper is to report on the calculation
of the leading electroweak corrections to the exclusive $bbH$
final state, meaning two $b$'s are detected. These leading
electroweak corrections are triggered by top-charged Goldstone
loops whereby, in effect, an external $b$ quark turns into a top.
This transition has a specific chiral structure whose dominant
part is given by the top mass or, in terms of couplings, to the
top Yukawa coupling. Considering that the latter is of the order
of the QCD coupling constant, the corrections might be large. In
fact, as we shall see, such type of transitions can trigger $g g
\ra b \bar b H$ even with {\em vanishing} $\lambda_{bbH}$ in which
case the process is generated solely at one-loop. We will quantify
the effect of such contributions.

This calculation belonging to the class of the $2 \ra 3$ processes
at the LHC, we will also cover some technical issues pertaining to
such calculations, like among other things the helicity amplitude
method we use and the occurrence  of vanishingly small (inverse)
Gram determinant. This determinant occurs when solving the system
of (linearly independent) tensor integrals in terms of the basis of
scalar integrals. The Gram matrix is constructed out of the scalar
products of the $(N-1)$ linearly independent momenta for a process
with $N$ external legs. The Gram determinant can vanish if the
momenta of the set are, for example, for some exceptional point in
phase space no longer linearly independent.
\\

\noindent In this paper we restrict ourselves to a Higgs mass in
the range preferred by the latest electroweak data\cite{range_MH},
in particular we confine the present study to $M_H<150$GeV.
Another reason for this choice is that, as we will briefly point
out, as the Higgs mass increases the loop induced cross section
increases and the loop integral starts showing instabilities. This
we have identified as a  Landau singularity which is a pinch
singularity of the loop integral. This has an interesting physical
origin: the rescattering of on-shell top quarks into $W$ bosons,
giving rise to $W$ boson fusion into Higgs. We leave this
important issue to another study though.

The plan of the paper is as follows.  In the next section we
present some general considerations concerning the properties and
structure of the calculation we have performed. We first briefly
review the tree-level LO amplitude and highlight
some symmetries of the helicity amplitudes. These symmetries are
maintained by QCD corrections but not by the electroweak
corrections we are studying. We then discuss the leading
approximation as given by the insertions of the
top-bottom-Goldstone Yukawa vertex. We classify the
contributions into three classes in the cases of the NLO
correction as well as the contributions that survive at one-loop
even for $\lambda_{bbH}=0$. In section~3 we give our
renormalisation scheme and discuss the inclusion of a top/Higgs
Yukawa enhanced contribution which can be considered as a
universal correction to Higgs processes related to the Higgs wave
function renormalisation and the renormalisation of the vacuum
expectation value. Section~4 gives an overview of some
calculational details in particular how the calculation is
organised.  Discussion on the loop integrals, the appearance of
spurious instabilities related to vanishing Gram determinants and
how these are cured depending on how the phase space integration
is carried out is also presented. We also discuss in this section
how we checked our results through ultraviolet finiteness and
gauge invariance. Section~5 presents and discusses  the numerical
results we find for the total cross section and various
distributions both at the level of the NLO electroweak correction
as well as the one-loop contribution that survives in the limit of
vanishing $\la_{bbH}$. Details about the helicity amplitude method
we used as well as the optimisation of the code are presented in
two appendices.

\section{General considerations}
Before discussing the details of the calculation it is educative
to expose some key features that appear when one considers the
electroweak corrections at one-loop compared to the structure we
have at tree-level or even the structure that emerges from QCD
loop calculations. In particular the helicity structure is quite
telling. So let us set our definition first. The process we
consider is
$g(p_1,\la_1)+g(p_2,\la_2)\rightarrow
b(p_3,\la_3)+\bar{b}(p_4,\la_4)+H(p_5)$. $\la_i=\pm$ with
$i=1,2,3,4$ are the helicities of the gluons, the bottom and
anti-bottom while $p_i$ are the momenta of particles. The
corresponding helicity amplitude will be denoted by ${\cal
A}(\la_1,\la_2;\la_3,\la_4)$.

\subsection{Leading order considerations}
\begin{figure}[h]
\begin{center}
\includegraphics[width=12cm]{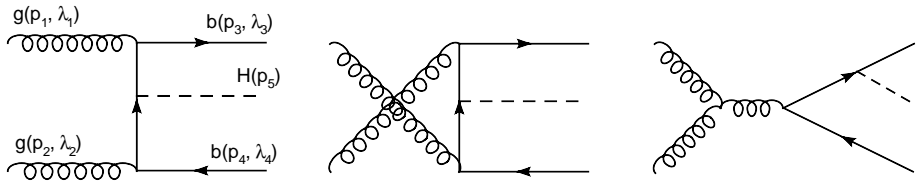}
\caption{\label{diag_gg_LO}{\em All the eight Feynman diagrams can
be obtained by inserting the Higgs line to all possible positions
in the bottom line.}}
\end{center}
\end{figure}

At tree-level,  see Fig. \ref{diag_gg_LO} for the contributing
diagrams, the Higgs can only attach to the $b$-quark and therefore
each diagram, and hence the total amplitude, is proportional to
the Higgs coupling to $b\bar b$, $\lambda_{bbH}$. Compared to the
gluon coupling this scalar coupling breaks chirality. These
features remain unchanged when we consider QCD corrections.
Moreover the QCD coupling and the Higgs coupling are parity
conserving which allows to relate the state with helicities
$(\la_1,\la_2;\la_3,\la_4)$ to the one with
$(-\la_1,-\la_2;-\la_3,-\la_4)$ therefore cutting by half the
number of helicity amplitudes to calculate. With our conventions
for the definition of the helicity states, see
Appendix~\ref{appendix-helicity},  parity conservation for the
tree-level helicity amplitude gives
\beqn
{\cal A}_0(-\la_1,-\la_2;-\la_3,-\la_4)=\la_3 \la_4 {\cal
A}_0(\la_1,\la_2;\la_3,\la_4)^\star.
\eeqn
This can be generalised at higher order in QCD with due care of
possible absorptive parts in taking complex conjugation.

The number of contributing helicity amplitudes can be reduced even
further at the leading order, in fact halved again, in the limit
where one neglects the mass of the $b$-quark that originates from
the $b$-quark spinors and therefore from the $b$ quark
propagators. We should in this case consider the $\lambda_{bbH}$
as an independent coupling, intimately related to the model of
symmetry breaking. In this case chirality and helicity arguments
are the same, the $b$ and $\bar b$ must have opposite helicities
for the leading order amplitudes and hence only ${\cal
A}_0(\la_1,\la_2;\la,-\la)$ remain non zero. In this limit, this
means that only a string containing an even number of Dirac
$\gamma$ matrices, which we will label in general as $\Gamma^{\rm
even}$ as opposed to $\Gamma^{\rm odd}$ for a string with an odd
number of $\gamma$'s, can contribute.

\noi In the general case and reinstating the $b$ mass, we may write the  helicity amplitudes  as
\bea \ali&=&\bar u(\lambda_3) \left( \Gamma^{\rm
even}_{\la_1,\la_2} + \Gamma^{\rm odd}_{\la_1,\la_2} \right)
v(\la_4) \crn &=&\delta_{\la_3, -\la_4} \left({\cal A}^{\rm
even}+m_b \tilde{{\cal A}}^{\rm odd} \right) \;+\;
\delta_{\la_3,\la_4} \left({\cal A}^{\rm odd}+m_b \tilde{{\cal
A}}^{\rm even} \right). \label{amp_form2}
\eea

The label $^{\rm even}$  in ${\cal A}^{\rm even}$ and
$\tilde{{\cal A}}^{\rm even}$ are the contributions of
$\Gamma^{\rm even}$ to the amplitude and likewise for $^{\rm
odd}$. This way of writing shows that $m_b$ originates from the
mass insertion coming from the massive spinors and are responsible
for chirality flip. In the limit $m_b \ra 0$, $\Gamma^{\rm
even}_{\la_1,\la_2}$ and $\Gamma^{\rm odd}_{\la_1,\la_2}$
contribute to different independent helicity amplitudes. In
general $\Gamma^{\rm even}$ and $\Gamma^{\rm odd}$  differ by a
(fermion) mass insertion. In fact $\Gamma^{\rm odd}$ is
proportional to a fermion mass insertion from a propagator. At
leading order the mass insertion is naturally $m_b$, such that
$\Gamma^{\rm odd}$ is ${\cal O}(m_b)$. This shows that at leading
order, corrections from $m_b=0$ to the total cross section are of
order ${\cal O}(m_b^2)$. Of course there might be some enhancement
of the ${\cal O}(m_b^2)$ terms if one remembers that the cross
section can bring about terms of order  $m_b^2/{(p_{T}^{b})}^2$.
However,  in our calculation where we require the $b$'s to be
observed hence requiring a $p_T^b$ cut, the effect will be
minimal. With $m_b=4.62$GeV, the effect of neglecting $m_b$
is that the cross section is increased by $3.7\%$ for
$|\textbf{p}_T^{b,\bar{b}}|>20$GeV and $1.1\%$ for
$|\textbf{p}_T^{b,\bar{b}}|>50$GeV. At one-loop, the chiral
structure of the weak interaction and the contribution of the top
change many of the characteristics that we have just discussed for
the tree-level.

\subsection{New electroweak  Yukawa-type contributions, novel characteristics}
\begin{figure}[h]
\begin{center}
\includegraphics[width=9cm]{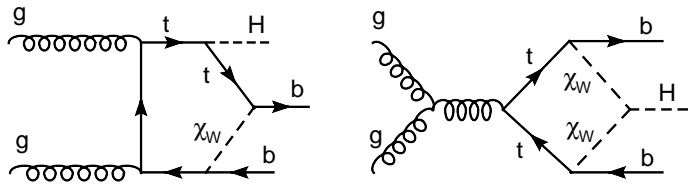}
\caption{\label{diag_gg_ew}{\em Sample of one-loop diagrams
related to the Yukawa interaction in the SM. $\chi_W$ represents
the charged Goldstone boson.}}
\end{center}
\end{figure}
\begin{figure}[htbp]
\begin{center}
\includegraphics[width=14cm]{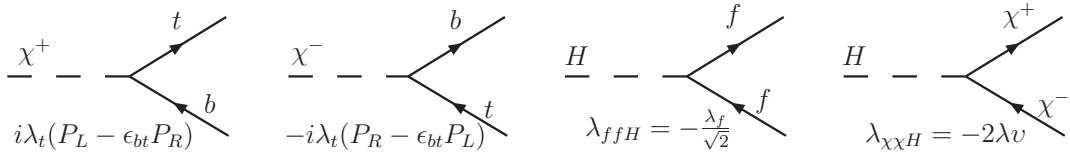}
\caption{\label{vertex_NLO}{\it Relevant vertices appearing at one
loop.  $\varepsilon_{bt}=\la_b/\la_t$, $v$ is the vacuum
expectation value and $\la$ is the Higgs self-coupling, related to
the Higgs mass in the Standard Model. $P_{L,R}=(1\mp\gamma_5)/2$}}
\end{center}
\end{figure}

Indeed, look at the two contributions arising from the one loop
electroweak corrections given in Fig.~\ref{diag_gg_ew}. Now the
Higgs can attach to the top or to the $W$. Therefore these
contributions do not vanish in the limit $\lambda_{bbH}=0$.
Because now the fermion loop is a top loop, the mass insertion in
what we called $\Gamma^{\rm odd}$ is proportional to the top mass
and is not negligible. In fact the diagrams in
Fig.~\ref{diag_gg_ew} show the charged Goldstone boson in the
loop. The latter triggers a $ t \ra b \chi_W$ transition whose
dominant coupling is proportional to the Yukawa coupling of the
top. We will in fact be working in the approximation of keeping
only the Yukawa couplings. This reduces the number of diagrams and
if working in the Feynman gauge as we do in this computation, only
the Goldstone contributions survive. The neutral Goldstone bosons
can only contribute corrections of order $\lambda_b^2$. We will
neglect these ${\cal O}(\la_b^2)$ contributions at the amplitude
level. However the order ${\cal O}(\la_b)$ corrections will be kept.
All the corrections are then triggered by $t \ra b
\chi_W$ and apart from the QCD $ g \ra b \bar b$ vertex, only the
Yukawa vertices shown in Fig.~\ref{vertex_NLO} below are needed to
build up the full set of electroweak corrections.

Note that in models outside the Standard Model, the Higgs coupling
to the fermion $f$, $\la_{ffH}$, can involve other parameters
beside the corresponding Yukawa coupling $\lambda_f$. The Higgs
coupling to the charged Goldstone involves the Higgs self-coupling
or Yukawa coupling of the Higgs, $\la=M_H^2/2 v^2$ proportional to
the square of the Higgs mass. The latter can be large for large
Higgs masses. These considerations allow to classify the
contributions into three gauge invariant classes.

\subsection{Three classes of diagrams and the chiral structure at one-loop}
\label{section_3classes}
\begin{figure}[hbt]
\begin{center}
\includegraphics[width=16cm]{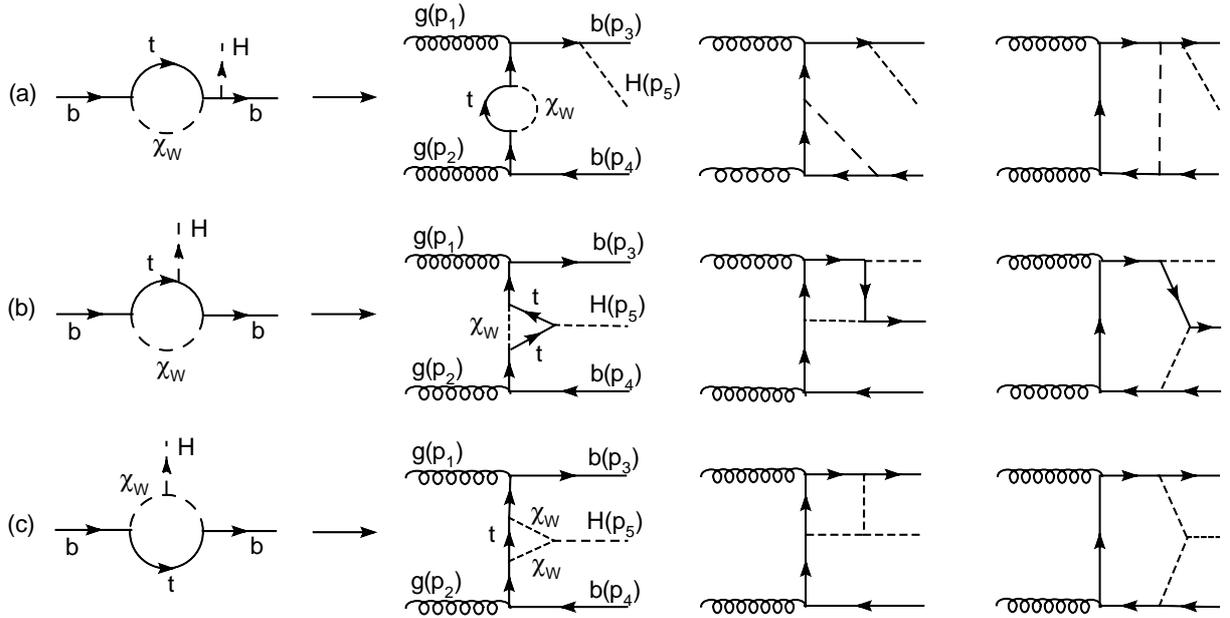}
\caption{\label{diag_3group}{\em All the diagrams in each groups
can be obtained by inserting the two gluon lines or one triple
gluon vertex (not shown) to all possible positions in the generic
bottom line, which is the first diagram on the left. We have checked the number of diagrams through Grace-loop\cite{grace}.}}
\end{center}
\end{figure}
All the one-loop diagrams are classified into three gauge
invariant groups as displayed in Fig.~\ref{diag_3group}. The Higgs
couples to the bottom quark in the first group (Fig.
\ref{diag_3group}a), to the top quark in the second group (Fig.
\ref{diag_3group}b) and to the charged Goldstone boson in the
third group (Fig. \ref{diag_3group}c). As shown in
Fig.~\ref{diag_3group} each class can be efficiently reconstructed
from the one-loop vertex $b \bar b H$, depending on which leg one
attaches the Higgs, by then grafting the gluons in all possible
ways. We have also checked explicitly that each class with its
counterterms, see below, constitutes a QCD gauge invariant subset.
Note that these three contributions depend on different
combinations of independent couplings and therefore constitute
independent sets.

The chiral structure $t \ra b \chi_W$ impacts directly on the
structure of the helicity amplitudes at one-loop. The split of
each contribution according to $\Gamma^{\rm even}$ and
$\Gamma^{\rm odd}$, see Eq.~\ref{amp_form2} will turn out to be
useful and will indicate which helicity amplitude can be enhanced
by which Yukawa coupling at one-loop. We show only one example in
class $(b)$ of Fig.~\ref{diag_3group}. It is straight forward to
carry the same analysis for all other diagrams. We choose the
first diagram in group (b) in Fig. \ref{diag_3group}. For clarity
we will here take $m_b=0$, we have already shown how $m_b$
insertions are taken into account, see Eq.~\ref{amp_form2}.
Leaving aside the colour part which can always be factorised out
(see Appendix~\ref{optimisation}) and the strong coupling
constant, we write explicitly the contribution of this diagram as

\bea
{\cal A}_{b1}(\la_1,\la_2;\la_3,\la_4)
=\la_{ttH}\la_t^2\bar{u}(\la_3,p_3)\slashep(\la_1,p_1)\fr{\slashpbar_{13}}{\bar{p}_{13}^2}C_{b1}\fr{\slashpbar_{24}}{\bar{p}_{24}^2}
\slashep(\la_2,p_2)u(\la_4,p_4) \label{amp_vertex}.
\eea
$C_{b1}$ is the Yukawa vertex correction. In $D$-dimension, with
$q$ the integration variable, the momenta as defined in
Fig.~\ref{diag_gg_LO} with $p_{ij}=p_i+p_j$ and $\bar{p}_{ij}=
p_j-p_i$ we have

\bea
C_{b1}=\int\fr{d^Dq}{(2\pi)^Di}\fr{(P_R-\varepsilon_{bt}P_L)(m_t+\slashq+\slashpbar_{13})(m_t+\slashq-\slashpbar_{24})
(P_L-\varepsilon_{bt}P_R)
}{(M_W^2-q^2)[m_t^2-(q+\bar{p}_{13})^2][m_t^2-(q-\bar{p}_{24})^2]}.\label{cb1}
\eea

The numerator of the integrand of (\ref{cb1}), neglecting terms of
${\cal O}(\la_b^2)$, can be re-arranged such as
\bea
{\cal A}_{b1}(\la_1,\la_2;\la_3,\la_4) \stackrel{\rm
numerator}{\longrightarrow} \underbrace{-\varepsilon_{bt}
\left(m_t^2+(\slashq+\slashpbar_{13})(\slashq-\slashpbar_{24})\right)}_{\Gamma^{\rm
even}} + \underbrace{m_t P_R \left(
2\slashq+\slashpbar_{13}-\slashpbar_{24}\right)}_{\Gamma^{\rm
odd}}. \label{C_numer}
\eea
This shows explicitly that ${\Gamma^{\rm odd}}$ structures with a
specific chirality, $P_R$, can indeed be generated. They do not
vanish as $\la_{bbh} \ra 0$. The even one-loop structures on the
other hand are ${\cal O}(\la_b)$. The structure in class $(c)$,
Higgs radiation off the charged Goldstones, is the same. For class
$(a)$, radiation off the $b$-quark, the structure of the
correction is different, the odd part is suppressed and receives
an  ${\cal O}(\la_b)$ correction. To summarise, with $m_b=0$,
 making explicit the Yukawa couplings and the chiral structure if any, for example $P_R$, that characterise
each class and comparing to the leading order, one has
\begin{center}
\begin{tabular}{|c|c|c|}
\cline{2-3}
\multicolumn{1}{c|}{ }   & ${\Gamma^{\rm even}}$ & ${\Gamma^{\rm odd}}$ \\
  \hline
  tree-level &$\la_{bbH}$ & 0 \\
  $(a)$ & $\la_t^2 \la_{bbH}$&  $\la_{b}\la_t \la_{bbH}$\\
  $(b)$ & $\la_{b} \la_{t} \la_{ttH}$ &  $\la_t^2 \la_{ttH}$, ($P_R$)\\
  $(c)$ & $\la_{b} \la_{t} \la_{\chi \chi H}$  & $\la_t^2 \la_{\chi \chi H}$, ($P_R$) \\
  \hline
\end{tabular}
\end{center}

We have kept $\la_{ffH}$ and $\la_f$ separate to show how the
structures may change in the MSSM for example and also why just by
inspecting the couplings we can differentiate between the three
classes. We clearly see that  all one-loop $\Gamma^{\rm even}$
contributions vanish in the limit $\la_{b}=0$ or $\la_{bbH}=0$. On
the other hand this is not the case for the one-loop $\Gamma^{\rm
odd}$ contribution belonging to class $(b)$ and $(c)$. However for
these contributions to interfere with the tree-level LO
contribution requires a chirality flip through a $m_b$ insertion.
Therefore in the SM for example, the NLO cross section is
necessarily of order $m_b^2$, like the LO, with corrections
proportional to the top Yukawa coupling for example. On the other
hand, in the limit of $\la_{bbH}=0$, the tree level vanishes
but $g g \ra b \bar b H$ still goes with an amplitude of order
$g_s^2 \la_t^2 \la_{ttH}$ or $g_s^2\la_t^2 \la_{\chi \chi H}$. For
$\la_{bbH} \neq 0$ these contributions should be considered as
part of the NNLO ``corrections" however they do not vanish in the
limit $m_b \ra 0$ (or $\la_{bbH}=0$) while the tree level does. These
contributions can be important and we will therefore study their
effects. For these  contributions at the ``NNLO" we can set
$m_b=0$.

The classification in terms of structures as we have done makes
clear also that the novel one-loop induced ${\Gamma^{\rm odd}}$
contributions must be ultraviolet finite. This is not necessarily
the case of the ${\Gamma^{\rm even}}$  structures where
counterterms to the tree-level structures are needed through
renormalisation to which we now turn.

\section{Renormalisation}\label{sec:renorm}
We use an on-shell (OS) renormalisation scheme exactly along the
lines described in\cite{grace}. Ultraviolet divergences are
regularised through dimensional regularisation. In our
approximation we only need to renormalise the vertices $b \bar b
g$ and $b \bar b H$ as well as the bottom mass, $m_b$. For the $b
\bar b g$ vertex, from the point of view of the corrections we are
carrying,  only wave function renormalisation for the $b/\bar b$
field is required: $\stackrel{(-)}{b} \kern -.600em {_{L,R}} \ra
(1+ \delta Z_{b_{L,R}}^{1/2}) \stackrel{(-)}{b} \kern -.600em
{_{L,R}}$. $\delta Z_{b_{L,R}}^{1/2}$ can be taken real,
see\cite{grace}. The counterterm to $m_b$, $\delta m_b$ and the
wave function renormalisation for the $b/\bar b$ are set by
imposing the usual conditions for pole position and residue on the
renormalised bottom propagator. In terms of the self-energy
correction $\Sigma_{bb}(q)$ with momentum $q$\cite{grace}:
\bea
\Sigma(q^2)=
K_1+K_\gamma\slashq+K_{5\gamma}\slashq\gamma_5.
\eea
This translates into
\bea
\delta m_b&=&Re\Big(m_b K_\gamma(m_b^2)+K_1(m_b^2)\Big),\crn \delta
Z_{b_L}^{1/2}&=&
\fr{1}{2}Re\Big(K_{5\gamma}(m_b^2)-K_\gamma(m_b^2)\Big)-m_b\fr{d}{dq^2}\Big(m_bReK_\gamma(q^2)+ReK_1(q^2)\Big)\/|_{q^2=m_b^2},\crn
\delta
Z_{b_R}^{1/2}&=&-\fr{1}{2}Re\Big(K_{5\gamma}(m_b^2)+K_\gamma(m_b^2)\Big)
-m_b\fr{d}{dq^2}\Big(m_bReK_\gamma(q^2)+ReK_1(q^2)\Big)\/|_{q^2=m_b^2}.\label{counter_vertex}
\eea
We calculate the coefficients $K_{1,\gamma,5\gamma}$ of the
bottom self-energy in the same spirit we calculate the other
one-loop corrections, {\it i.e.} only through the $ t \ra b
\chi_W$ transition, see the first diagram of class $(a)$ in
Fig.~\ref{diag_3group}. We get,

\bea
K_1(q^2)&=&-\fr{\la_t^2}{16\pi^2} \;\Big(
C_{UV}-F_0(m_t,M_W,q^2)\Big),\crn
K_{\gamma}(q^2)&=&-K_{5\gamma}(q^2)= \fr{\la_t^2}{64\pi^2} \;
\Big( C_{UV}-2 F_1(m_t,M_W,q^2) \Big)\crn {\rm with} \;\;
C_{UV}&=&\fr{1}{\ep}-\gamma_E+\ln 4\pi\,\,,D=4-2\ep,\crn
F_n(m_1,m_2,q^2)&=&\int_0^1dxx^n\ln
\Big((1-x)m_1^2+xm_2^2-x(1-x)q^2 \Big). \label{KFG}
\eea

The reason we get $K_{\gamma}(q^2)=-K_{5\gamma}(q^2)$ is due to
the particular chiral structure of the $t \ra b \chi_W$ loop
insertion. In particular for $m_b=0$, one recovers that these
corrections only contribute to $\delta Z_{b_L}^{1/2}$ and not
$\delta Z_{b_R}^{1/2}$.

The counterterms needed to renormalise the $b \bar b H$ vertex are
$\delta m_b, \delta Z_{b_{L,R}}^{1/2}$ as well as the Higgs wave
function renormalisation $\delta Z_{H}^{1/2}$ and the counterterm
to the vacuum expectation value, $\upsilon$, $\delta \upsilon$.
Indeed we have $\delta_{bbH}=\la_{bbH}(\fr{\delta m_b}{m_b}
+\delta Z_{b_L}^{1/2} +\delta Z_{b_R}^{1/2} +(\delta
Z_H^{1/2}-\delta\upsilon))$. The $ t \ra b \chi_W$ loop insertion
does not contribute to $\delta Z_{H}^{1/2}$ (which originates from
the Higgs self energy two-point function) nor to $\delta
\upsilon$, the renormalisation of the vacuum expectation value. On
the other hand $(\delta Z_H^{1/2}-\delta\upsilon)$ can be seen as
a universal correction to Higgs production processes. We will
include this correction as it has potentially large contributions
scaling like $\la_t^2$ and $\la$ which fall into the category of
the corrections we are seeking. Within the calculation we have
performed this means that the combination $(\delta
Z_H^{1/2}-\delta\upsilon)$ must be finite. Indeed, we find
\bea
\delta Z_H^{1/2}&=&-\fr{1}{8\pi^2}Re\left\{\fr{3\la_{t}^2}{4}
\Big(C_{UV}-F_0(m_t,m_t,M_H^2)\right.\crn
&-& M_H^2G_0(m_t,m_t,M_H^2)+4m_t^2G_0(m_t,m_t,M_H^2)\Big)
\crn &-&\left.\fr{\la}{4}\Big(9 G_0(M_H,M_H,M_H^2)+2
G_0(M_W,M_W,M_H^2)+G_0(M_Z,M_Z,M_H^2)\Big)\right\},\nn \eea
\bea
\delta\upsilon &=&-\fr{1}{8\pi^2}Re\left\{\fr{3\la_t^2}{4}
\Big(C_{UV}-2F_1(m_b,m_t,M_W^2)\Big) \right.\crn &-& \la
\Big(F_0(M_H,M_W,M_W^2)-F_1(M_H,M_W,M_W^2)-\fr{1}{2}\ln
M_H^2\Big)\crn
&-&\fr{c_W^2}{s_W^2}\Big(\fr{3\la_t^2}{4}\big(F_0(m_t,m_t,M_Z^2)-2F_1(m_b,m_t,M_W^2)\big)
+ \la \big(F_0(M_H,M_Z,M_Z^2) \crn
 &-&\left. F_1(M_H,M_Z,M_Z^2)   - F_0(M_H,M_W,M_W^2)+F_1(M_H,M_W,M_W^2)\big)\Big)\right\},\crn
& & \hspace*{-1cm}
G_n(m_1,m_2,q^2)=q^2\fr{d}{dq^2}F_n(m_1,m_2,q^2)=q^2\int_0^1dx\fr{-x^nx(1-x)}{(1-x)m_1^2+xm_2^2-x(1-x)q^2},\crn
\;\;\;\;\;\;
\eea
which shows that $(\delta Z_H^{1/2}-\delta\upsilon)$ is finite.
\\ \noi In the actual calculation, the counter term $\delta_{bbg}^\mu$
belongs to class $(a)$ in the classification of
Fig~\ref{diag_3group}. This makes class $(a)$ finite. The
counterterm we associate to class $(b)$ is the part of
$\delta_{bbH}$ from the $t \ra b \chi_W$ loops and therefore does
not include what we termed the universal Higgs correction, {\it
i.e} does not include the contribution $(\delta
Z_H^{1/2}-\delta\upsilon)$. This is sufficient to make class $(b)$
finite. In our approach $(c)$ is finite without the addition of a
counterterm. We will keep the $(\delta Z_H^{1/2}-\delta\upsilon)$
contribution separate from the  contributions in  classes
$(a),(b),(c)$. We will of course include it in the final result.

\section{Calculation details}
We have written two independent codes. In the first one we set
$m_b=0$ in all propagators and other spinors that emerge from the
helicity formalism we follow. In this limit, the helicity
formalism is very much simplified and the expression quite
compact. This code is in fact subdivided in two separate
sub-codes. One sub-code is generated for the ``even" part
(constituted by the $\Gamma^{\rm even}$ contributions, see
~Eq.~\ref{amp_form2}) and the other by the ``odd" part. We also
generate a completely independent code for the case $m_b \neq 0$
where in particular we use the helicity formalism with massive
fermions. Details of the helicity formalism that we use are given
in Appendix~\ref{appendix-helicity}.

The steps that go into writing these codes are the following. In
the first stage, we use Form\cite{form} to generate expressions
for the tree level and one loop helicity amplitudes. Each helicity
amplitude is written in terms of Lorentz invariants, scalar spinor
functions $(A,B,C)_{\la_i\la_j}$ defined in
Appendix~\ref{appendix-helicity} and the
Passarino-Veltman\cite{pass_velt} tensor functions $T^{N}_{M}$ for
a tensor of rank $M$ for $N$-point function. We have also sought
to write the contribution of each amplitude as a product of
different structures or blocks that reappear for different graphs
and contributions. For example colour factorisation is
implemented, this further allows to rearrange the amplitude into
an Abelian part and a non-Abelian part which will not interfere
with each other at the matrix element squared level. The helicity
information is contained in a set of basic blocks for further
optimisation. Another set of blocks pertains to the loop integrals
and other elements. The factorisation of the full
amplitude in terms of independent building blocks is easily
processed within Form. These building blocks can still consist of
long algebraic expressions which can be efficiently abbreviated
into compact variables with the help of a Perl script which also
allows to convert the output of Form into the Fortran code ready
for a numerical evaluation. More details on the connection between
Form and Fortran as well as the optimisation we implemented in the
codes can be found in Appendix~\ref{optimisation}.

\subsection{Loop integrals, Gram determinants and phase space integrals}
\label{sec:gram}
The highest rank $M$ of the Passarino-Veltman
tensor functions $T^{N}_{M}$ with $M \leq N$ that we encounter in
our calculation is $M=4$ and is associated to a pentagon graph,
$N=5$. We use the library {\tt LoopTools}\cite{looptools} to
calculate all the tensorial one loop integrals as well as the
scalar integrals, this means that we leave it completely to {\tt
LoopTools} to perform the reduction of the tensor integrals to the
basis of the scalar integrals. In order to obtain the cross
section one needs to perform the phase-space integration and
convolution over the gluon distribution function (GDF), $g(x,Q)$
with $Q$ representing the factorisation scale. We have
\bea
\sigma(p p\to b\bar{b}H
)&=&\fr{1}{256}\int_0^1dx_1g(x_1,Q)\int_0^1dx_2g(x_2,Q)\crn
&\times &\fr{1}{\hat
F}\int\fr{d^3\textbf{p}_3}{2e_3}\fr{d^3\textbf{p}_4}{2e_4}\fr{d^3\textbf{p}_5}{2e_5}|{\cal
A}(gg\to b\bar{b}H)|^2\delta^4(p_1+p_2-p_3-p_4-p_5)
\,,\crn\label{sigma_pp}
\eea
where $\fr{1}{256}=\fr{1}{4}\times\fr{1}{8}\times\fr{1}{8}$ is the
spin and colour average factor and  the flux factor is
$1/\hat{F}=1/\Big(2\pi)^52\hat{s}\Big)$ with $\hat{s}=x_1x_2s\ge
(2m_b+M_H)^2$.\\
\noi The integration over the three body phase space and momentum
fractions of the two initial gluons is done by using two
``integrators": {\tt BASES}\cite{bases} and {\tt
DADMUL}\cite{dadmul}. {\tt BASES} is a Monte Carlo that uses the
importance sampling technique while  {\tt DADMUL} is based on the
adaptive quadrature algorithm. The use of two different phase
space integration routines helps control the accuracy of the
results and helps detect possible instabilities. In fact some
numerical instabilities in the phase space integration do occur
when we use {\tt DADMUL} but not when we use {\tt BASES} which
gives very stable results with small integration error, typically
$0.08\%$ for $10^5$ Monte Carlo points. For the range of Higgs
masses we are studying in this paper, the instabilities that are
detected with {\tt DADMUL} were identified as spurious
singularities having to do with vanishing Gram determinants for
the three and four point tensorial functions calculated in
LoopTools by using the Passarino-Veltman reduction
method\footnote{The reduction of the five point function using the
method of  Denner and Dittmaier~\cite{denner_5p,looptools_5p}
which avoids the Gram determinant at this stage as implemented in
{\tt LoopTools} gives very stable results.}. Because this problem
always happens at the boundary of phase space, we can avoid it by
imposing appropriate kinematic cuts in the final state. In our
calculation, almost all zero Gram determinants disappear when we
apply the  cuts on the transverse momenta of the bottom quarks
relevant for our situation, see section~\ref{results} for the
choice of cuts. The remaining zero Gram determinants occur when
the two bottom quarks or one bottom quark and the Higgs are
produced in the same direction. Our solution, once identified as
spurious, was to discard these points by imposing some tiny cuts
on the polar, $\theta$, and relative azimuthal angles, $\phi$ of
the outgoing $b$-quarks, the value of the cuts is $\theta_{\rm
cut}^{b,\bar b}=\vert\sin\phi^{\bar b}\vert_{\rm cut}=10^{-6}$. {\tt DADMUL}
then produces the same result as {\tt BASES} within the
integration error.

\subsection{Checks on the results}
i) Ultraviolet finiteness: \\
\noi The final results must be ultraviolet (UV) finite. It means
that they should be independent of the parameter $C_{UV}$ defined
in Eq.~\ref{KFG}. In our code this parameter is treated as a
variable.The cancellation of $C_{UV}$ has been carefully checked
in our code. Upon varying the value of the parameter $C_{UV}$ from
$C_{UV}=0$ to $C_{UV}=10^{5}$, the results is stable within more
than 9 digits using double precision. This check makes sure that the divergent part of
the calculation is correct. The correctness of the finite part is
also well checked in our code by confirming that each helicity
configuration is QCD gauge invariant.\\
\noi ii) QCD gauge invariance: \\
\noi In the physical gauge we use, the
QCD gauge invariance reflects the fact that the gluon is massless
and has only two transverse polarisation components. In the
helicity formalism that we use, the polarisation vector of the
gluon of momentum $p$ and helicity $\la$ is constructed with the
help of a reference vector $q$, see Appendix~\ref{appendix-helicity} for details. 
The polarisation vector is
then labelled as $\eps^{\mu}(p,\la;q)$. A change of reference
vector from $q$ to $q^\prime$ amounts essentially to a gauge
transformation (up to a phase)
\bea
\eps^\mu(p,\la;q^\prime)=e^{i\phi(q^\prime,q)}\eps^\mu(p,\la;q)+\beta(q^\prime,q)p^\mu
.
\eea
QCD gauge invariance in our case amounts to independence of the
cross section in the choice of the reference vector, $q$. We have
carefully checked that the numerical result for the norm of each
helicity amplitude at various points in  phase space is
independent of the reference vectors say $q_{1,2}$ for gluon 1 and
2, up to 12 digits using double precision. By default, our
numerical evaluation is based on the use of $q_{1,2}=(p_2,p_1)$.
For the checks in the case of massive $b$ quarks the result with
the default choice $q_{1,2}=(p_2,p_1)$ is compared with a random
choice of $q_{1,2}$, keeping away from vectors with excessively
too small or too large components, see
Appendix~\ref{appendix-helicity} for more details.\\
\noi iii) As stated earlier, the result based on the use of  the
massive quark helicity amplitude are checked against those with
the independent code using the massless helicity amplitude by
setting the mass of the $b$ quark to zero. This is though just a
consistency check.\\
\noi iv) At the level of integration over phase space and density
functions we have used two integration routines and made sure that
we obtain the same result once we have properly dealt with the
spurious Gram determinant as we explained in
section~\ref{sec:gram}.\\
\noi v) Moreover, our tree level results have been successfully
checked against the results of CalcHEP\cite{calchep}.

\section{Results}
\subsection{Input parameters and kinematical cuts}
\label{results} Our input parameters are $\alpha(0)=1/137.03599911$,
$M_W=80.3766$GeV, $M_Z=91.1876$GeV, $\alpha_s(M_Z)=0.118$,
$m_b=4.62$GeV, $m_t=174.0$GeV with $s_W\equiv
\sqrt{1-M_W^2/M_Z^2}$. The CKM parameter $V_{tb}$ is set to be $1$.
We consider the case at the LHC where the center of
mass energy of the two initial protons is $\sqrt{s}=14$TeV.
Neglecting the small light quark initiated contribution, we use CTEQ6L\cite{cteq6}
for the GDF in the proton. The
factorisation scale for the GDF and energy scale for the strong
coupling constant are chosen to be $Q=M_Z$ for simplicity.

As has been done in previous analyses~\cite{dawson_bbH,LH03_bbh},
for the exclusive $b\bar{b}H$ final state, we require the outgoing
$b$ and $\bar{b}$ to have high transverse momenta
$|\textbf{p}_{T}^{b,\bar{b}}|\ge 20$GeV and pseudo-rapidity
$|\eta^{b,\bar{b}}|<2.5$. These kinematical cuts reduce the total
rate of the signal but also greatly reduce the QCD background. As
pointed in~\cite{dittmaier_bbH} these cuts also stabilise the
scale dependence of the QCD NLO corrections compared to the
case where no cut is applied. In the following, these kinematical
cuts are always applied unless otherwise stated.

Talking of the NLO QCD scale uncertainty and before presenting our
results, let us remind the reader of the size of the QCD
corrections. Taking a renormalisation/factorisation scale as we
take here at $M_Z$, the QCD corrections in a scheme where the
bottom Yukawa coupling is taken on-shell amount to $\sim -22\%$
for a Higgs mass of $120$GeV.

\subsection{NLO EW correction with $\la_{bbH}\neq 0$ }
\begin{figure}[hb]
\begin{center}
\mbox{\includegraphics[width=0.45\textwidth]{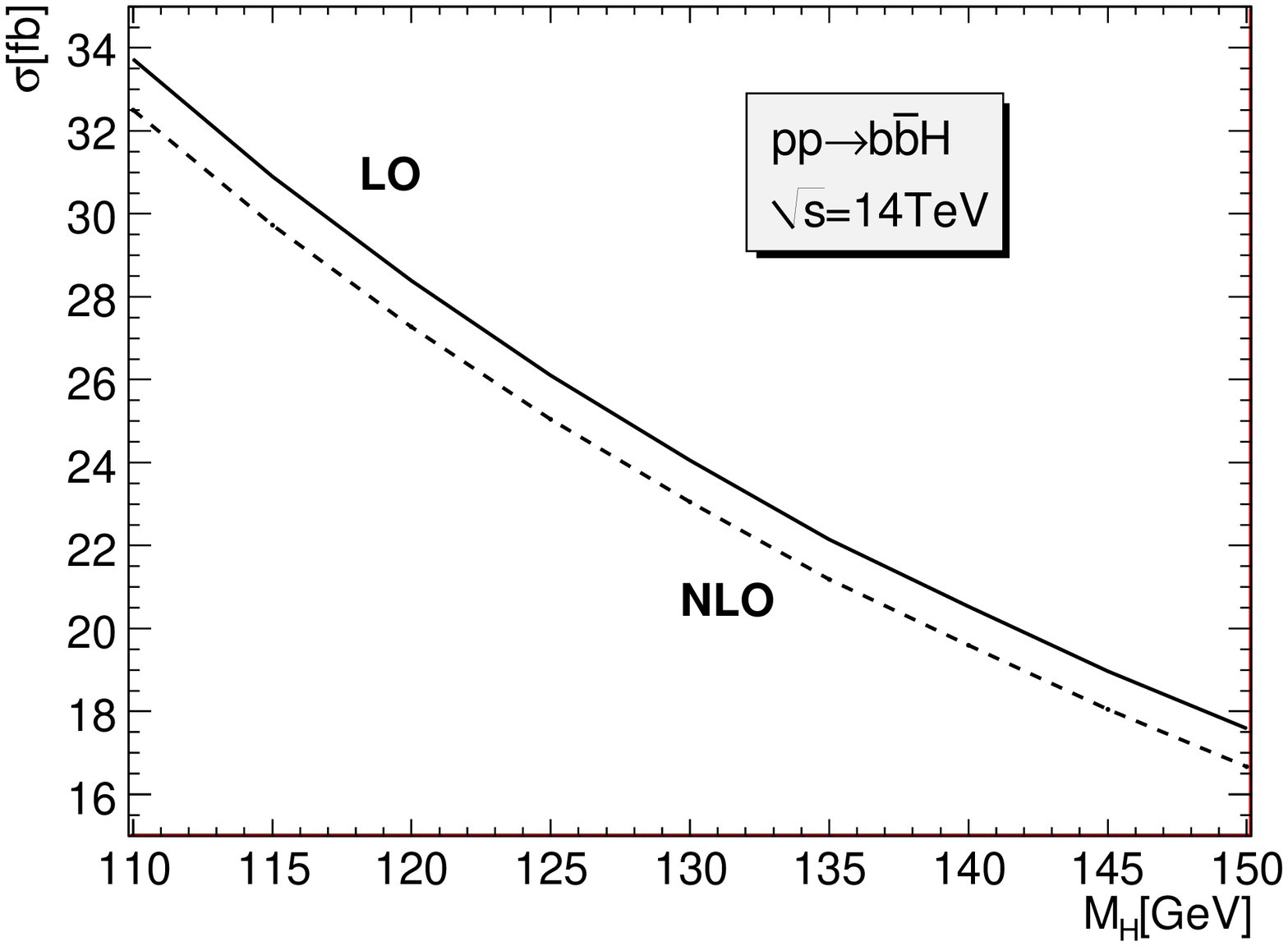}
\hspace*{0.075\textwidth}
\includegraphics[width=0.45\textwidth]{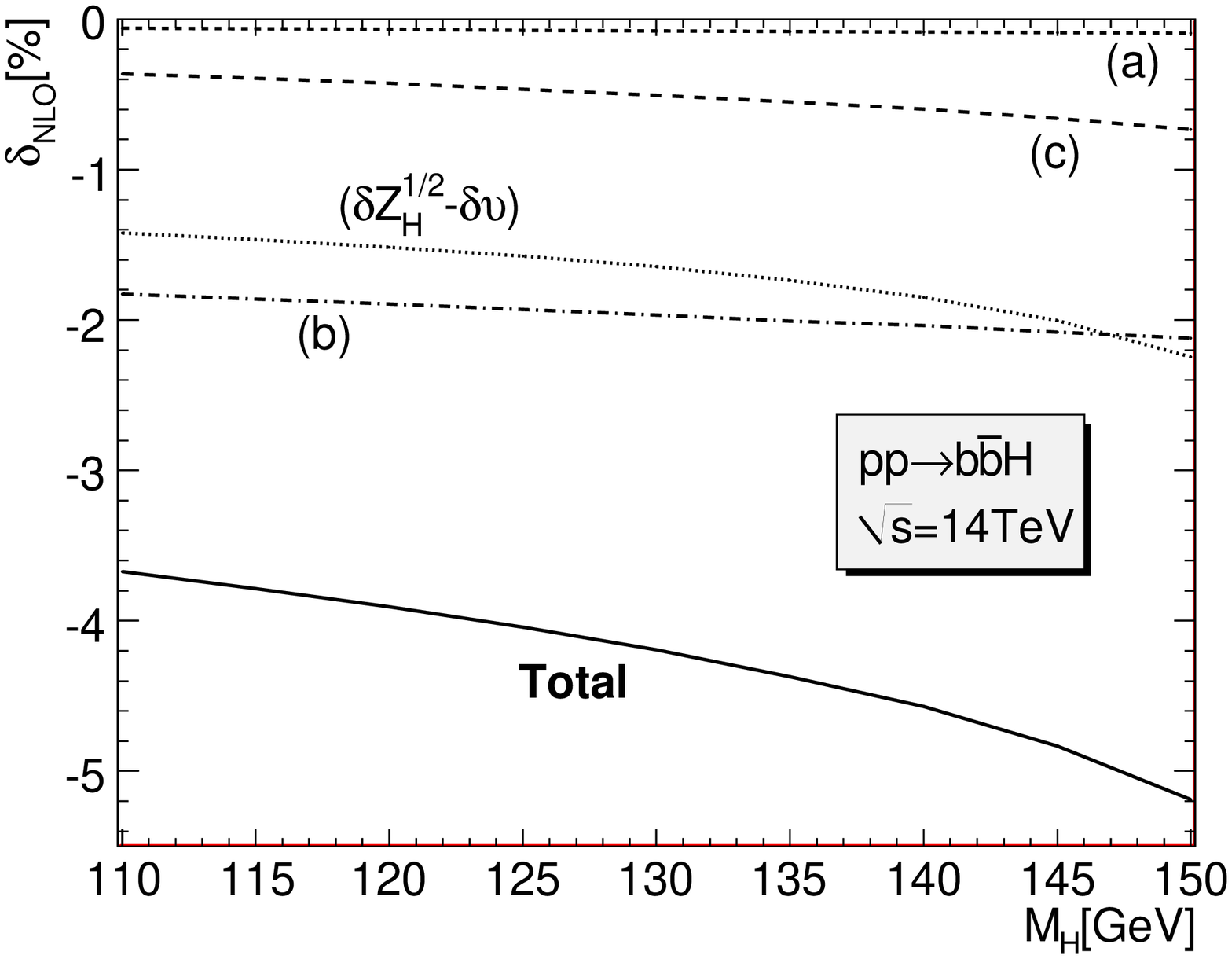}}
\caption{\label{p_LT_mH}{\em Left: the LO and NLO cross sections
as functions of $M_H$. Right: the relative NLO EW correction
normalized to tree level $\sigma_{LO}$. $(a)$, $(b)$, $(c)$ correspond to
the three classes of diagrams as displayed in Fig.
\ref{diag_3group} to which counterterms are added (see
section~\ref{sec:renorm}). $(\delta Z_{H}^{1/2}-\delta\upsilon)$
is the correction due to the  universal correction contained in
the renormalisation of the $b\bar bH$ vertex. ``Total'' refers to
the total electroweak correction, of Yukawa type, at one-loop.}}
\end{center}
\end{figure}

The cross sections with two high-$p_T$ bottom quarks at LO  and
NLO at the LHC are displayed in Fig. \ref{p_LT_mH} as a function
of the Higgs mass. The NLO EW correction reduces the cross section
by about $4\%$ to $5\%$ as the Higgs mass is varied from  $110$GeV
to $150$GeV. The first conclusion to draw is that this correction
is small if we compare it to the QCD correction or  even to the
QCD scale uncertainty. Considering that we have pointed to the
fact that the contributions could be grouped into three gauge
invariant classes that reflect the strengths of the Higgs coupling
to the $b$, the $t$ or its self-coupling, one can ask whether this
is the result of some cancellation. It turns out not to be the
case. All contributions are below $3\%$, see~Fig.~\ref{p_LT_mH}.
Class $(a)$ with a Higgs radiated from the bottom line is totally
negligible ranging from $-0.09\%$ to $-0.06\%$. We have failed in
finding a good reason for the smallness of this contribution
compared to the others. Those  due to the Higgs self-coupling are
below $1\%$. Radiation from the top contributes about $-2\%$ and
is of the same order as the contribution of the universal
correction. We had argued that the Yukawa corrections brought
about by the top might be large. It seems that the mass of the top
introduces also a large scale which can not be neglected compared
to the effective energy of the hard process even for LHC
energies.\\
\begin{figure}[htb]
\begin{center}
\mbox{\includegraphics[width=0.45\textwidth]{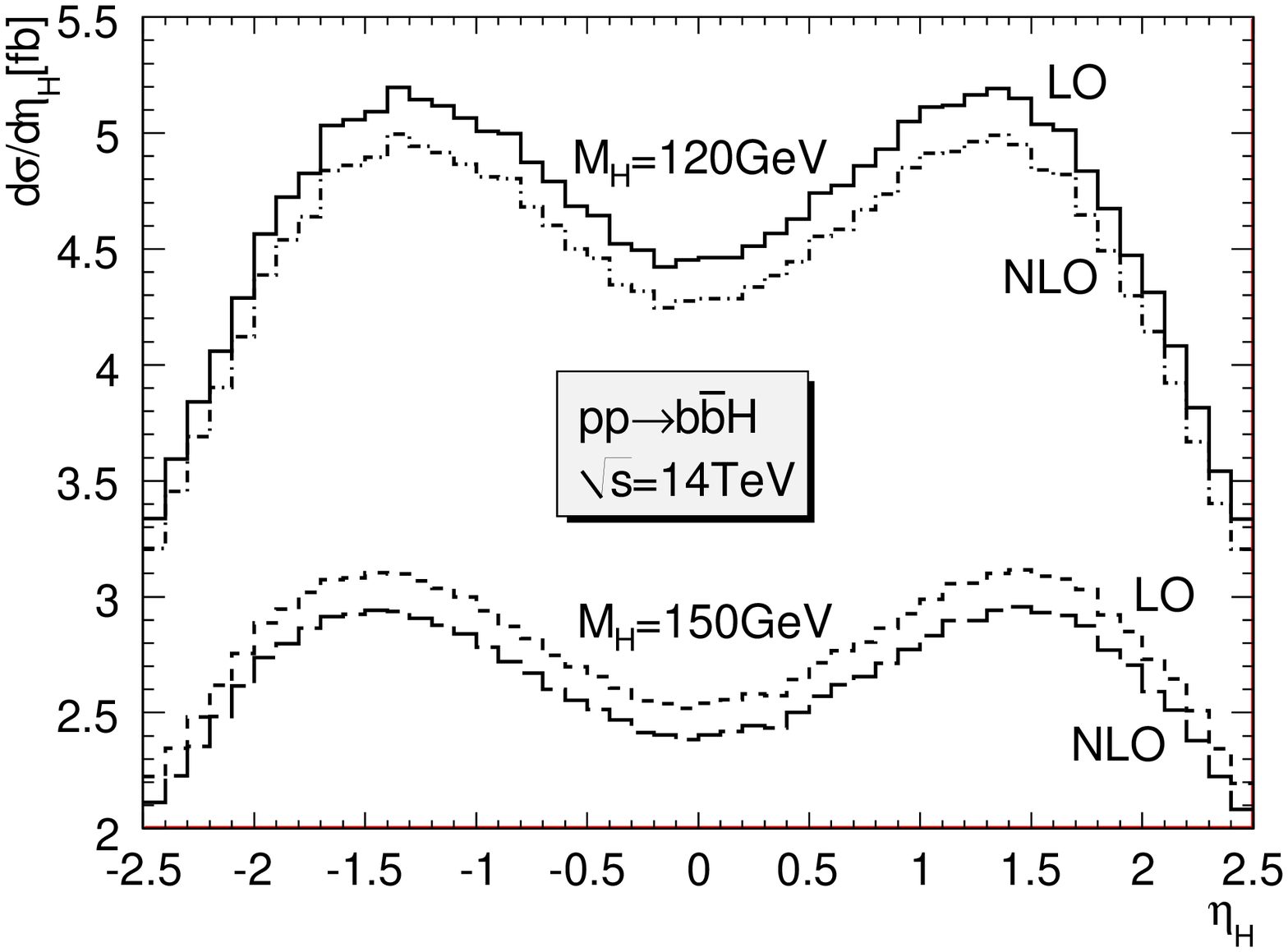}
\hspace*{0.075\textwidth}
\includegraphics[width=0.45\textwidth]{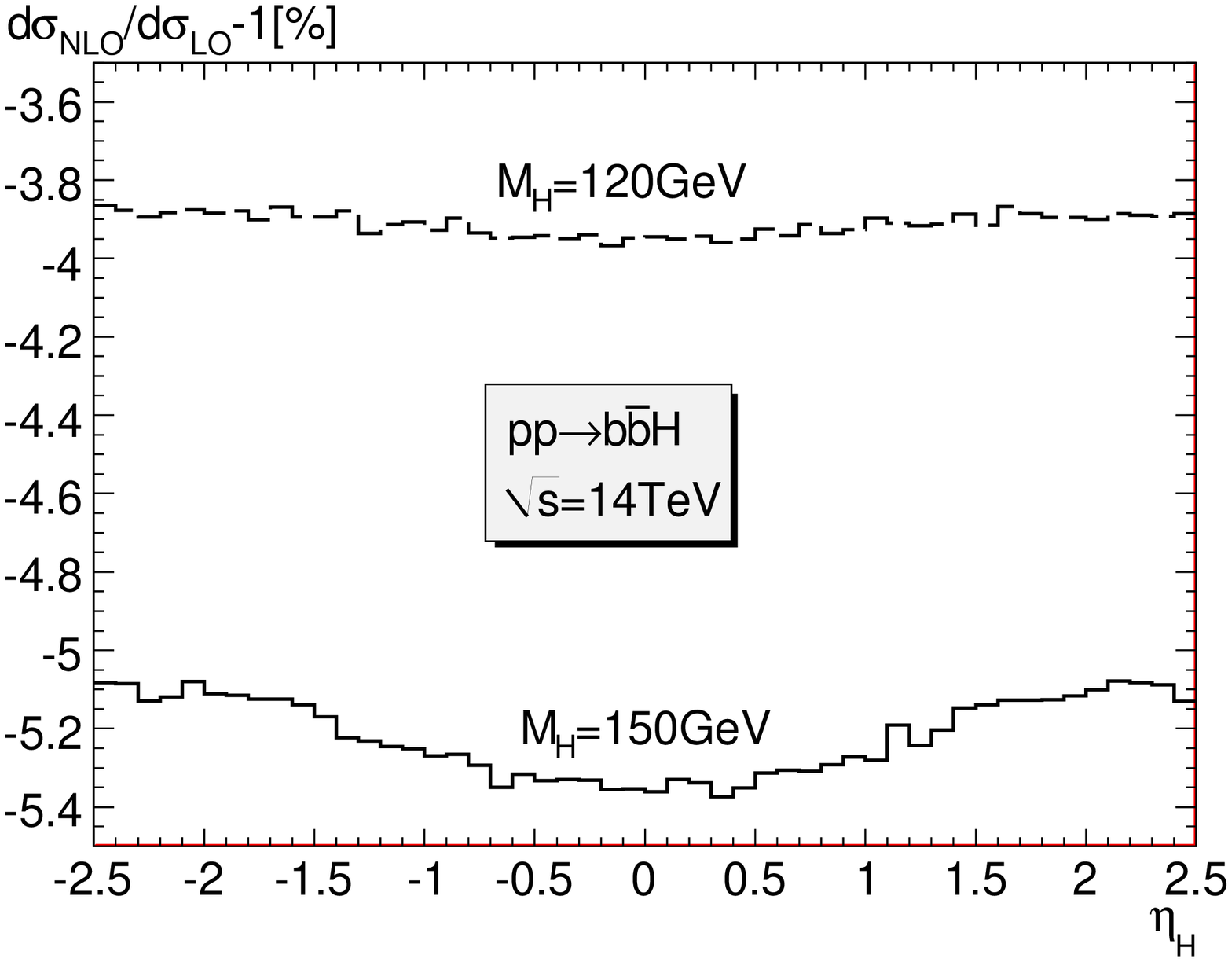}}
\mbox{\includegraphics[width=0.45\textwidth]{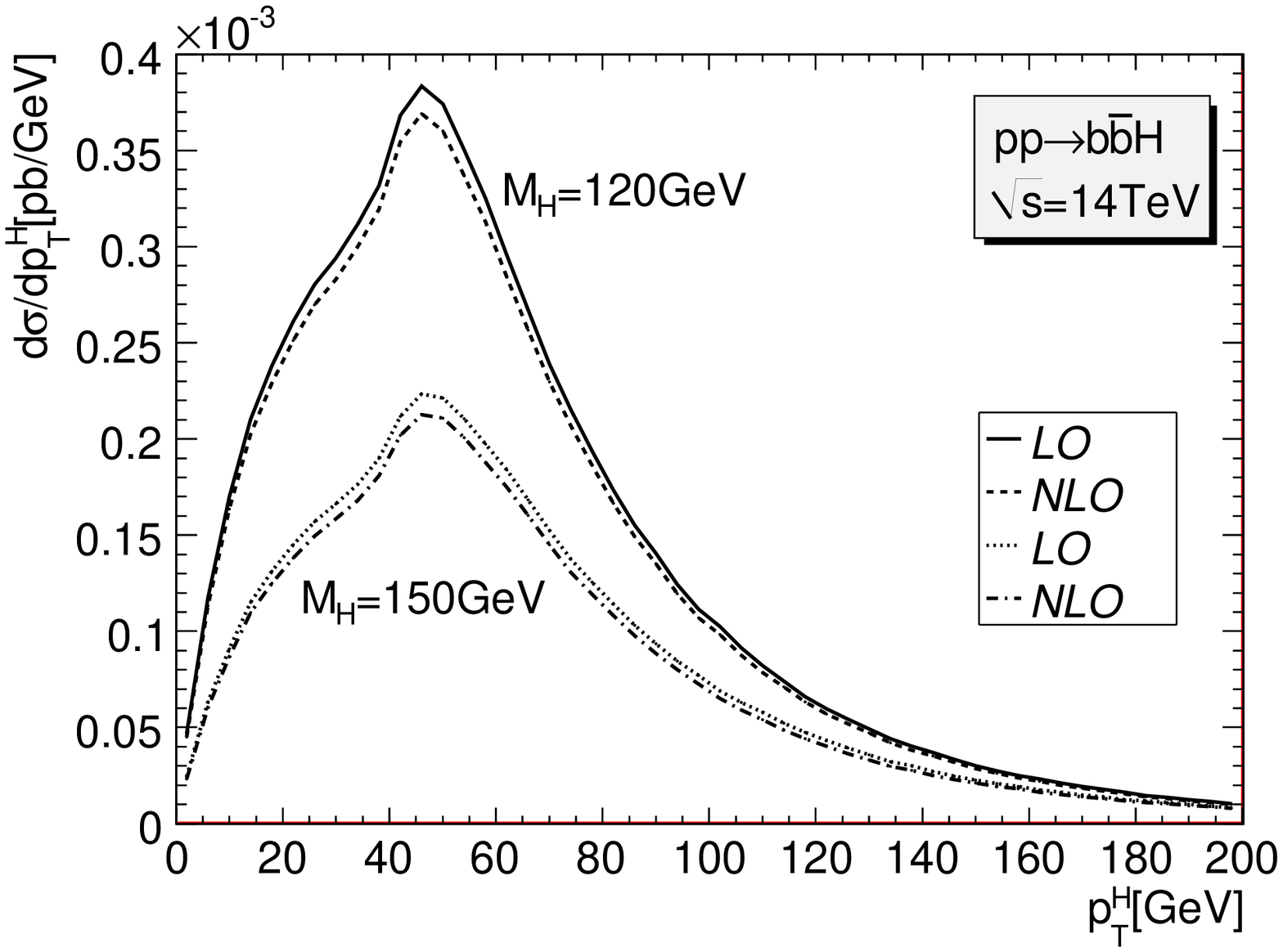}
\hspace*{0.075\textwidth}
\includegraphics[width=0.45\textwidth]{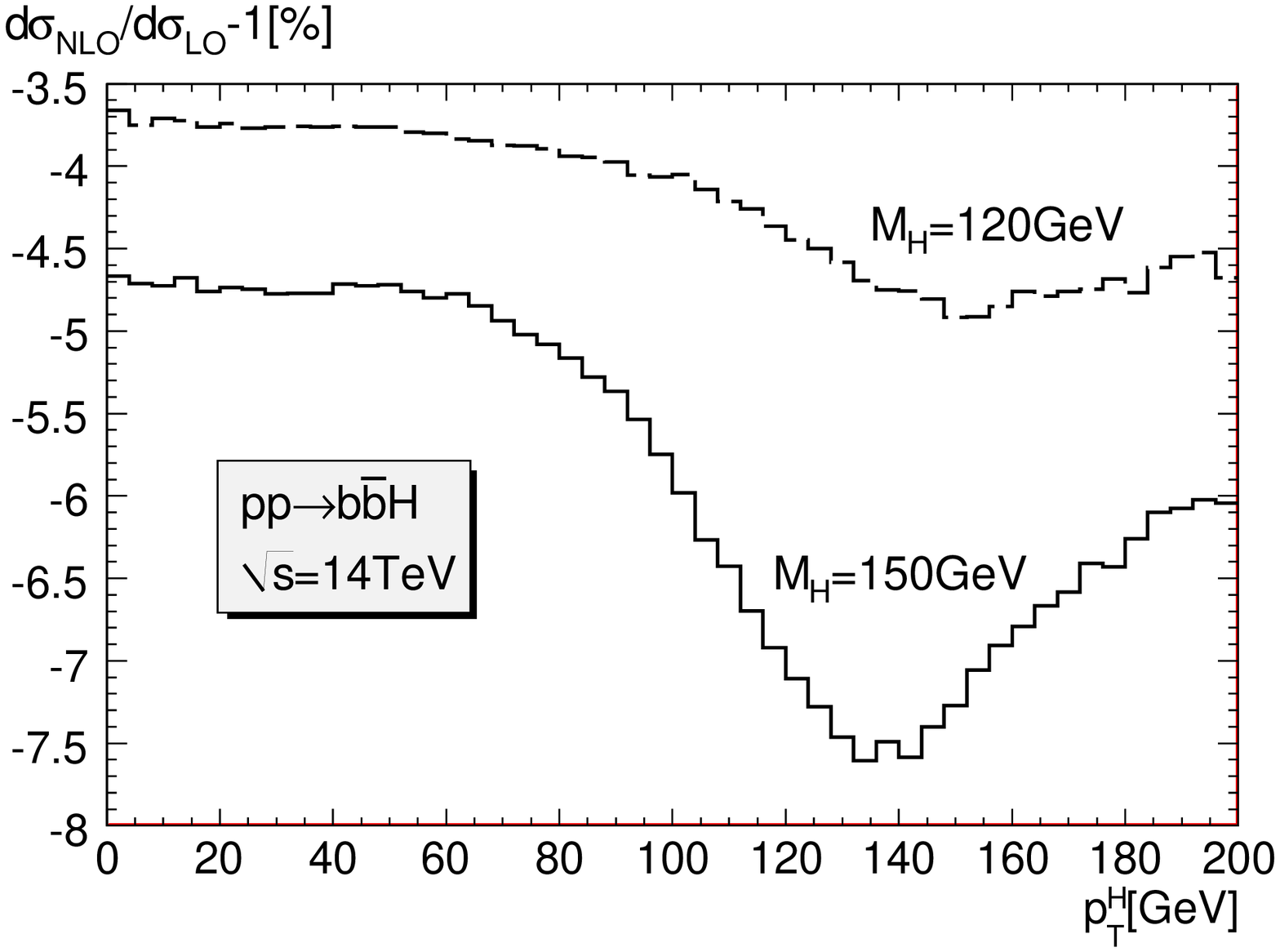}}
\caption{\label{fig:dist-nlo}{\em Effect of the NLO electroweak
corrections on the pseudo-rapidity and transverse momentum
distributions of the Higgs for $M_H=120,150$GeV. The relative
corrections $d\sigma_{NLO}/d\sigma_{LO}-1$ is also shown.}}
\end{center}
\end{figure}

The NLO corrections are spread rather uniformly on all the
distributions we have looked at. We have chosen to show in
Fig.~\ref{fig:dist-nlo} the effect on pseudo-rapidity and
transverse momentum distributions of the Higgs for two cases
$M_H=120$GeV and $M_H=150$GeV.  As  Fig.~ \ref{fig:dist-nlo} shows
the relative change in these two distributions is sensibly
constant especially for $M_H=120$GeV. For $M_H=150$GeV, the
corrections are largest for $p_T^H$ around $140$GeV, however this
is where the cross section is very small. A similar pattern, {\it
i.e.} a constant change in the distributions, is observed for the
bottom variables.

\subsection{EW correction in the limit of vanishing $\la_{bbH}$}
\begin{figure}[hbtp]
\begin{center}
\mbox{\includegraphics[width=0.45\textwidth]{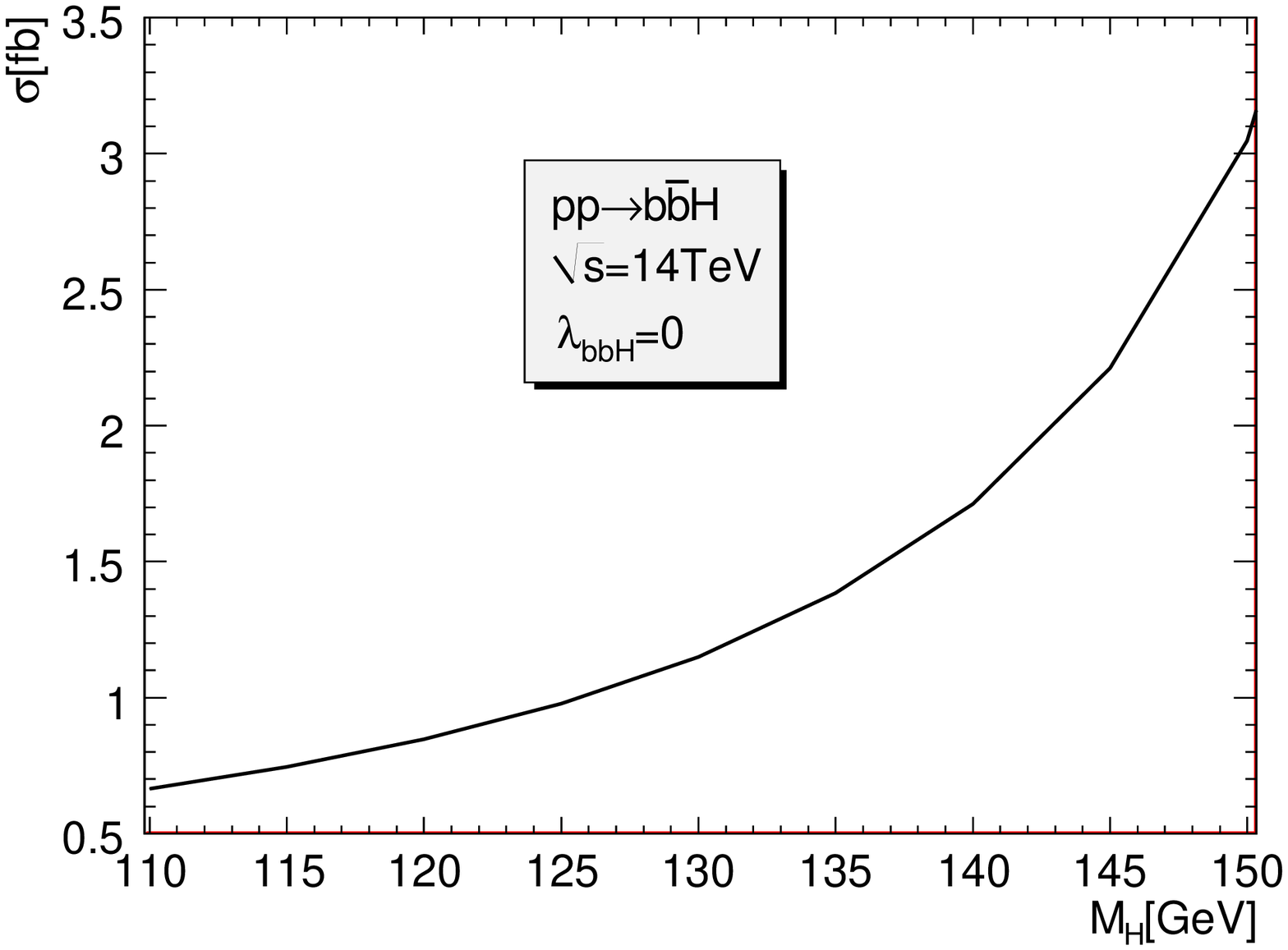}
\hspace*{0.075\textwidth}
\includegraphics[width=0.45\textwidth]{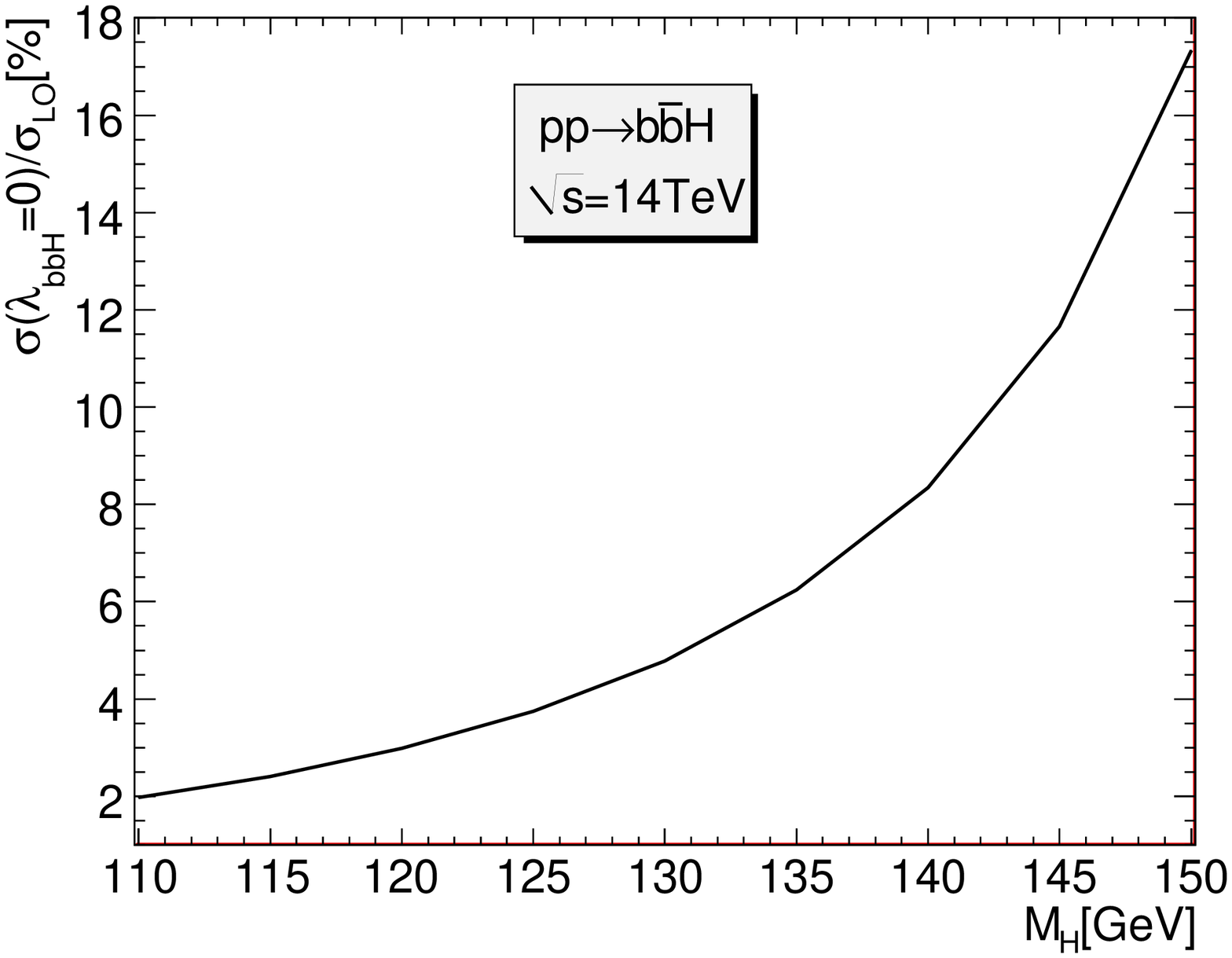}}
\caption{\label{p_LL_mH}{\em The one-loop induced cross section as
a function of $M_H$ in the limit of vanishing bottom-Higgs Yukawa
coupling. The right panel shows the percentage contribution of
this contribution relative to the tree level cross section calculated with
$\la_{bbH}\neq 0$.}}
\end{center}
\end{figure}

The cross section for $\la_{bbH}=0$ can be induced at one-loop
through the top loop. This ``NNLO" contribution rises rather
quickly as the Higgs mass increases even in the narrow range
$M_H=110-150$GeV as can be seen in Fig.~\ref{p_LL_mH}. Indeed
relative to the tree level,  the  cross section with $M_H=120$GeV amounts
to $3\%$ while for $M_H=150$GeV it has increased to as much as
$17\%$. Going past $M_H\geq 2M_W$ we encounter a Landau
singularity\cite{landau_pole} (a pinch singularity in the loop
integral) from diagrams like the one depicted in
Fig.~\ref{diag_gg_ew} (right) with the Higgs being attached to the $W$'s
or their Goldstone counterpart. It corresponds to a situation
where all particles in the loop are resonating and can be
interpreted as the production and decay of the tops into
(longitudinal) $W$'s with the later fusing to produce the Higgs.
This leading Landau singularity is not integrable, at the level of
the loop amplitude squared and must be regulated by the
introduction of  a width for the unstable particles. We leave this
issue together with a general discussion of Landau singularities
in such situations to another publication.
\begin{figure}[hp]
\begin{center}
\mbox{\includegraphics[width=0.45\textwidth]{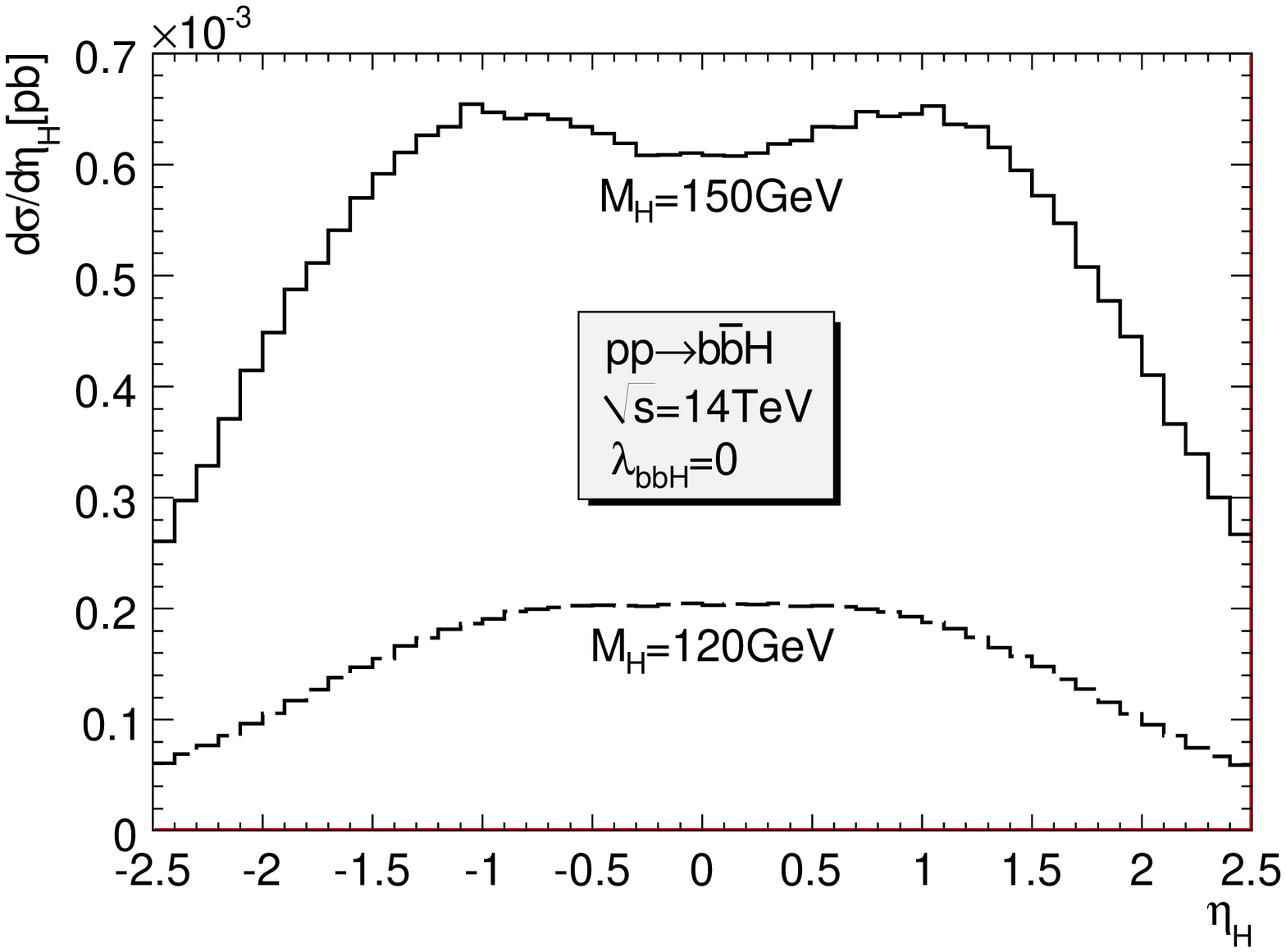}
\hspace*{0.075\textwidth}
\includegraphics[width=0.45\textwidth]{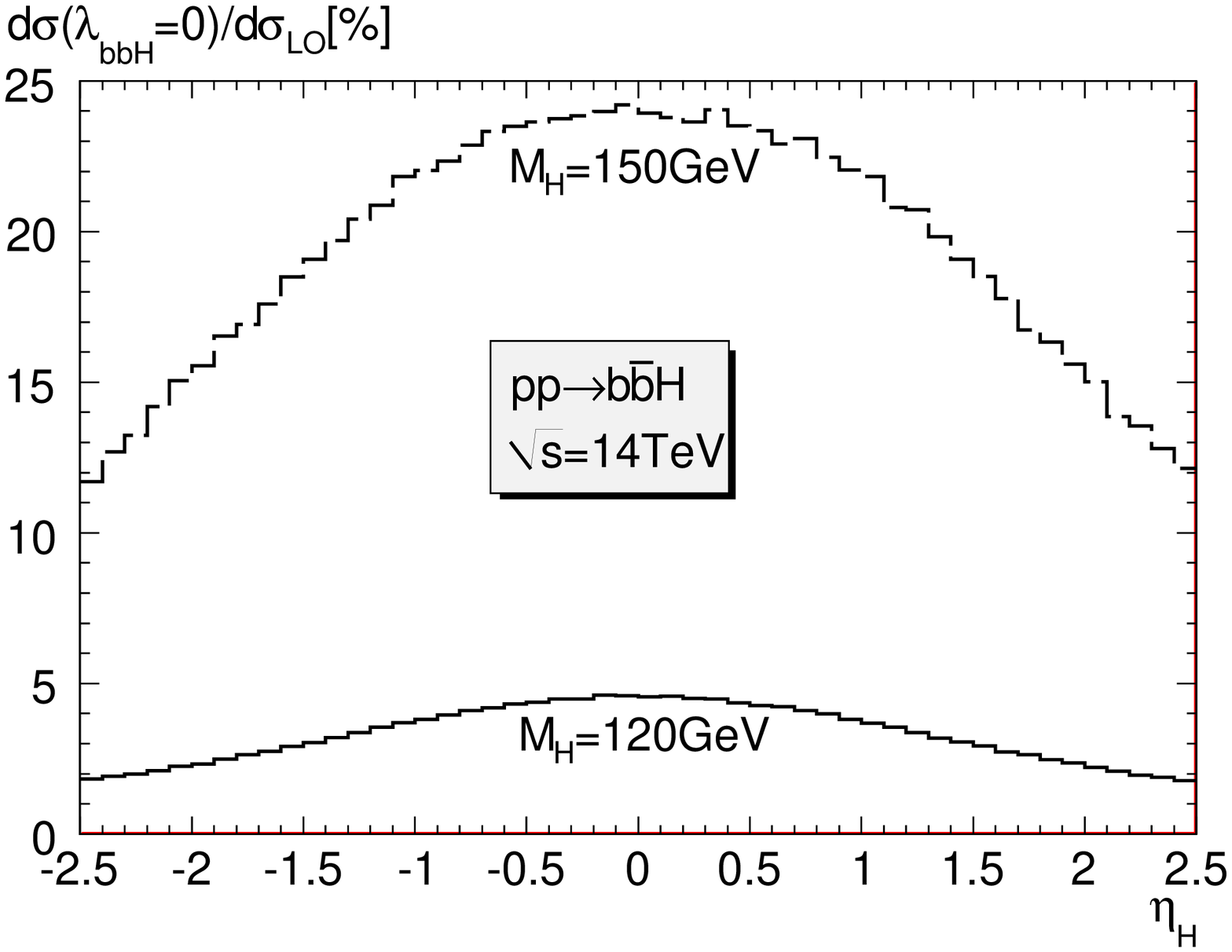}}
\mbox{\includegraphics[width=0.45\textwidth]{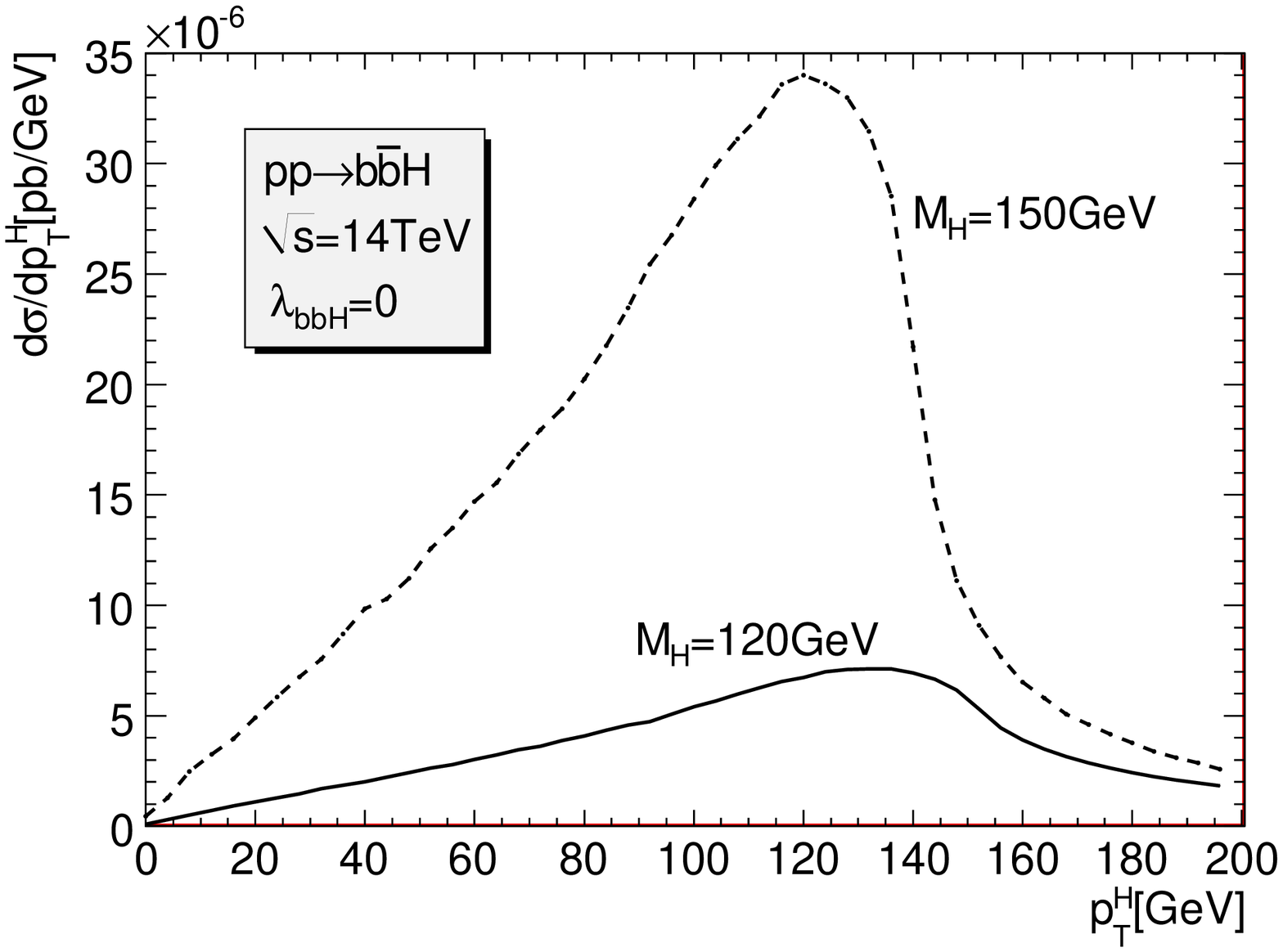}
\hspace*{0.075\textwidth}
\includegraphics[width=0.45\textwidth]{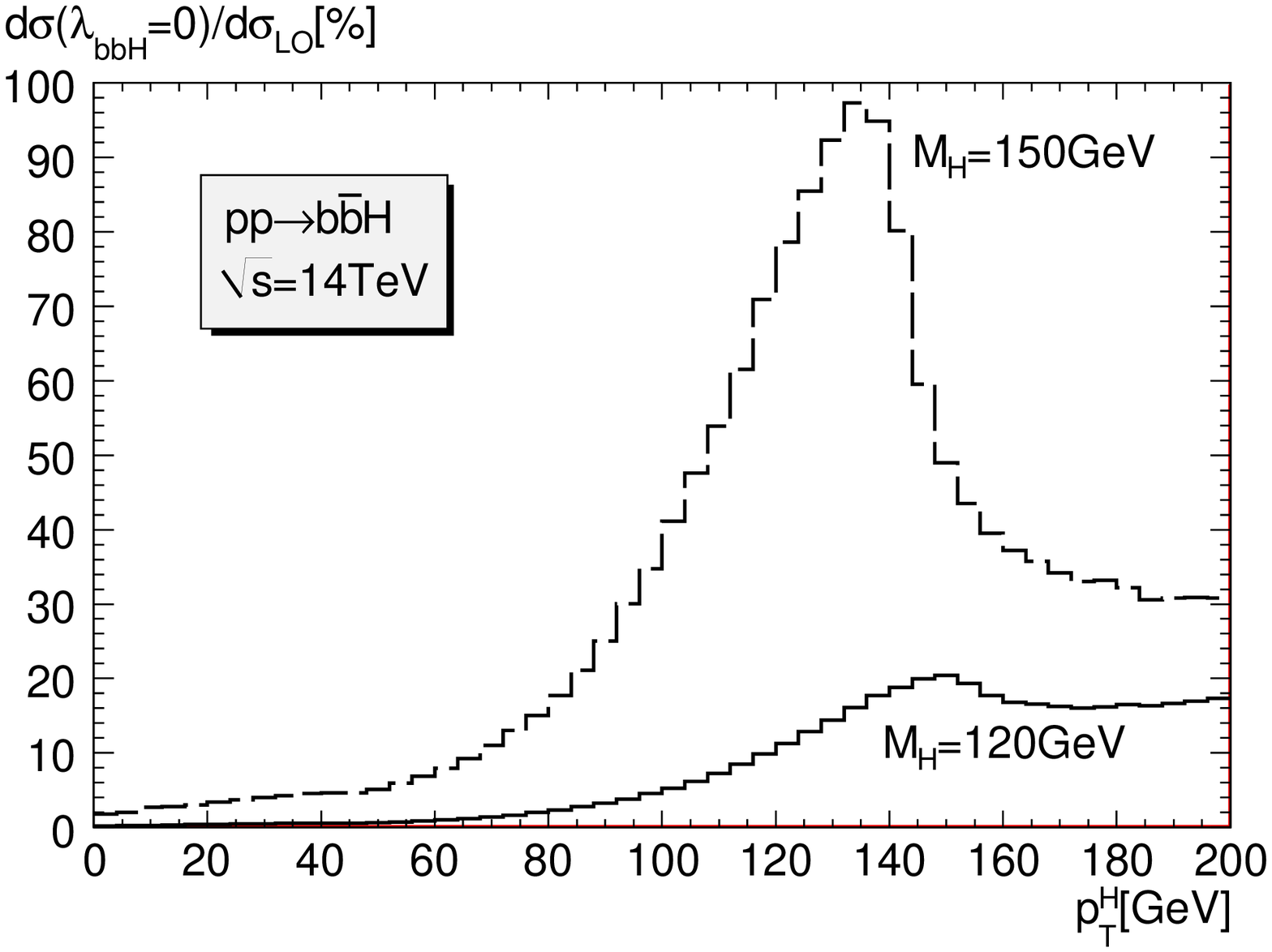}}
\mbox{\includegraphics[width=0.45\textwidth]{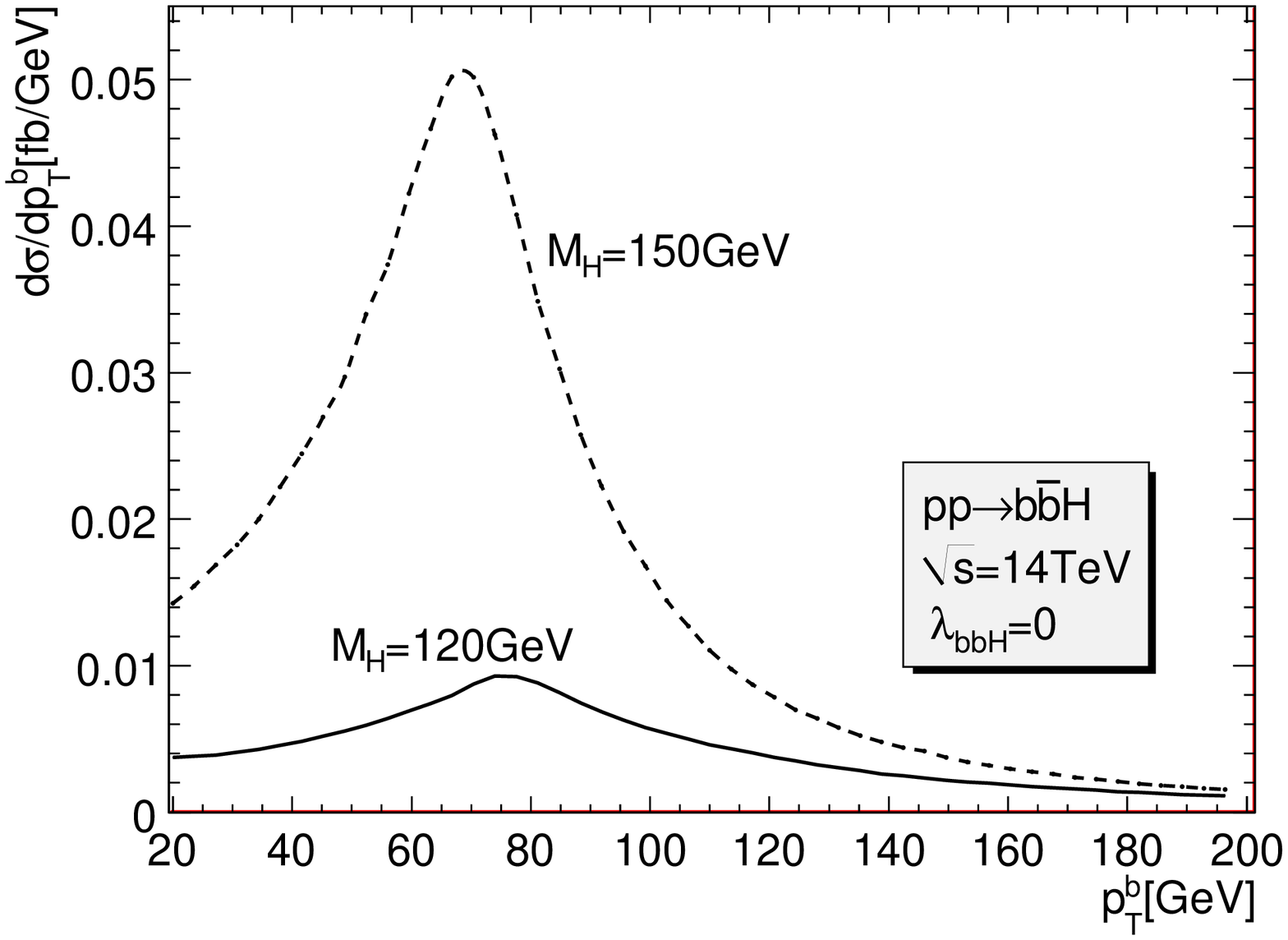}
\hspace*{0.075\textwidth}
\includegraphics[width=0.45\textwidth]{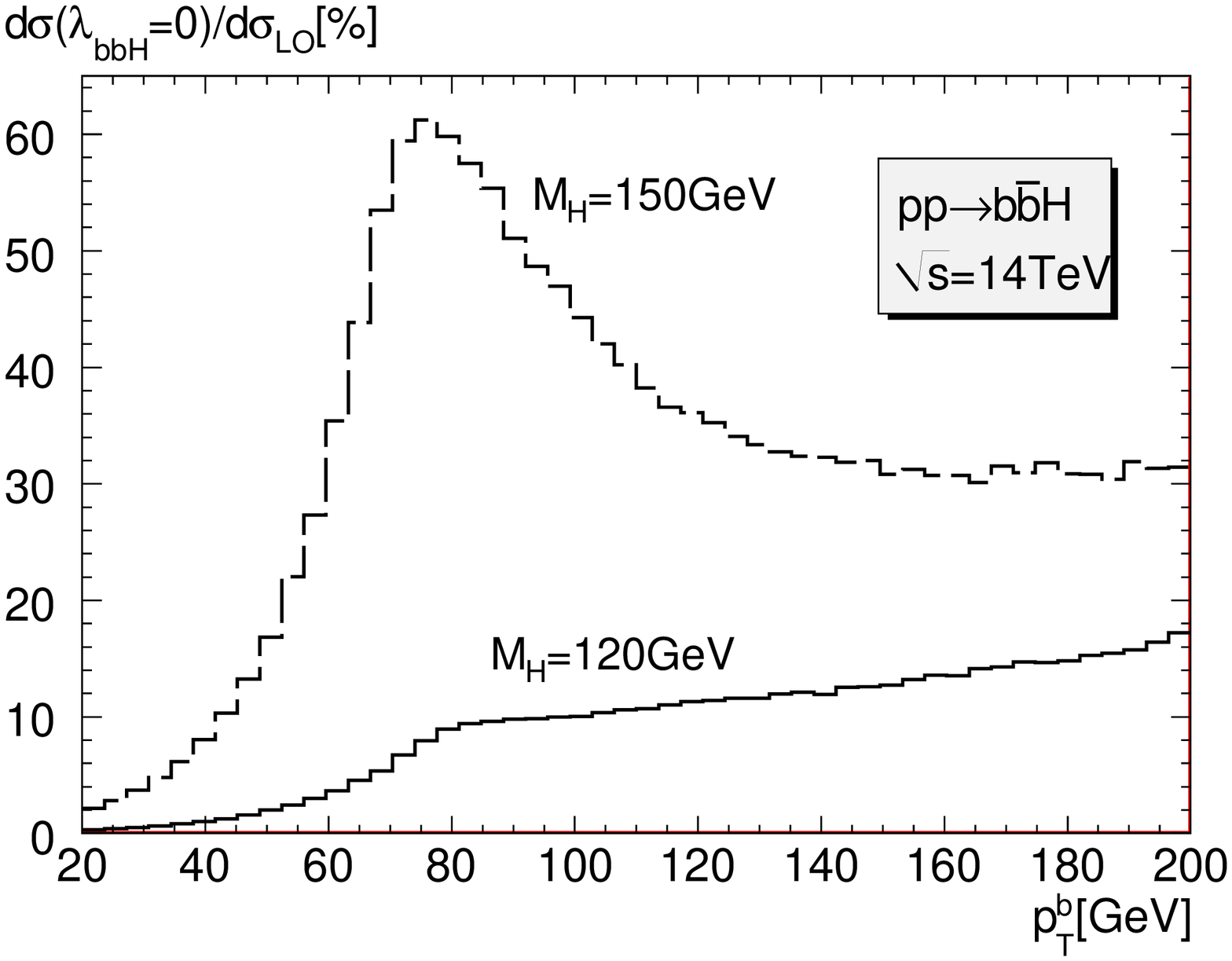}}
\caption{\label{p_LL_eta}{\em The pseudo-rapidity of the Higgs and
transverse momentum distributions of the Higgs and the bottom for
$M_H=120,150$GeV arising from the purely one-loop contribution in
the limit of vanishing LO ($\la_{bbH}=0$). Its relative percentage
contribution $d\sigma(\lambda_{bbH}=0)/d\sigma_{LO}$ is also
shown}}
\end{center}
\end{figure}

Fig. \ref{p_LL_eta}  shows the pseudo-rapidity and transverse
momentum distributions of the Higgs as well as the the $p_T$ of
the bottom for two cases $M_H=120$GeV and $M_H=150$GeV in the
limit of vanishing bottom-Higgs Yukawa coupling. These
distributions are significantly different from the ones we
observed at tree-level (and with the electroweak NLO corrections),
see Fig.~\ref{fig:dist-nlo}. The Higgs prefers being produced at
high value of transverse momentum, about $130$GeV. In the case of
a Higgs with $M_H=150$GeV this contribution can  significantly
distort the shape of the $p_T^H$ distribution for hight $p_T^H$
with a "correction" of more than $70\%$ over a rather large range.
The distribution in the $p_T$ of the bottom is also very telling.
The new contributions do not produce the bottom preferentially
with low $p_T^b$ as the case of the LO contribution.

\section{Conclusions}
\begin{figure}[htb]
\begin{center}
\includegraphics[width=8cm]{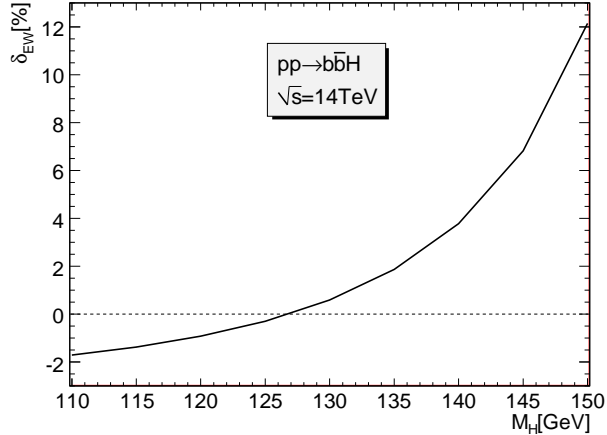}
\caption{\label{p_LandLL_mH}{\em
$\delta_{EW}=\delta_{NLO}+\fr{\sigma(\la_{bbH}=0)}{\sigma_0}$ as a
function of $M_H$.}}
\end{center}
\end{figure}

We have calculated the EW radiative corrections triggered by the
Yukawa coupling of the top to the process $p p \to b\bar{b}H$ at
the LHC through gluon fusion in the Standard Model. This process
is triggered through Higgs radiation of the bottom quark with a
small coupling proportional to the mass of the bottom. Yet in
order to analyse this coupling, precision calculations that
include both the QCD and electroweak corrections are needed. In
this perspective, to identify the process  one needs to tag both
$b$-jets. Our calculation is therefore conducted in this
kinematical configuration. Inserting a top quark loop with a
Yukawa transition of the type $t \ra b \chi_W$, $\chi_W$ is the
charged Goldstone, allows now the Higgs  to be radiated from the
top or from the Goldstone boson. The latter coupling represents
the Higgs self-coupling and increases with the Higgs mass. The
former, the top Yukawa coupling, is also large. As a consequence,
the one-loop  amplitude $g g  \to b\bar{b}H$ no longer vanishes as
the Higgs coupling to $b$'s does, like what occurs  at leading
order. We find that in the limit of vanishing $\la_{bbH}$, the
one-loop induced electroweak process should be taken into account
for Higgs masses larger than $140$GeV or so. Indeed, though this
contribution is quite modest for a Higgs mass of $110$GeV it
increases quite rapidly as the Higgs mass increases, reaching
about $17\%$ of the leading order value, calculated with
$m_b=4.62$GeV, for $M_H=150$GeV. For these new corrections to
interfere with the leading order requires helicity flip. Therefore
at next-to-leading order in the Yukawa electroweak corrections,
all corrections involve either a bottom mass insertion or a bottom
Yukawa coupling. At the end the total Yukawa electroweak NLO
contribution brings in a correction which is within the range
$-4\%$ to $-5\%$ for Higgs masses in the range $110{\rm GeV} < M_H
< 150$GeV. They are therefore negligible compared to the NLO QCD
correction and even the remaining QCD scale uncertainty. This
modest effect translates also as an uniform rescaling of the
distributions in the most interesting kinematical variables we
have looked at (pseudo-rapidities and $p_T$ of both $b$-quarks and
the Higgs). This is not the case of the one-loop induced
contributions which survive in the limit of $m_b \ra 0$ (and
$\la_{bbH} \ra 0$). Here the distributions for the Higgs masses
where the corrections for the total cross section is large are
drastically different from the LO distributions. A summary for the
corrections including the NLO with $\la_{bbH} \neq 0 $ and the
part of the NNLO counted as loop induced in the limit $\la_{bbH}
\ra 0$ is shown in Fig.~\ref{p_LandLL_mH}.

The analysis we have performed in this paper does not cover Higgs
masses over $150$GeV and rests within the range of Higgs masses
preferred by indirect precision measurements. In fact as the
threshold for $ H \ra WW$ opens up, important phenomena take
place. Foremost a Landau singularity, or a pinch singularity in
some loop integrals, develops. This corresponds the rescattering
of on-shell top quarks that decay to on-shell $W$ with Higgs
production via $WW$ fusion. We leave this important issue to a
forthcoming publication especially that the identification and
handling of such singularities can be applied to other processes.
In our case the singularity can be tamed by introducing the width
of the unstable particles. At NLO, for $M_H=2M_W$ for example, the
wave function renormalisation of the Higgs, which involves the
derivative of the two-point function Higgs self-energy, diverges.
This can also be regulated by including the width of the $W$, see
for example\cite{Kniehl-wfrh}.

There is another contribution which does not vanish for
vanishing $\lambda_{bbH}$ and which contributes to $gg \ra b \bar
b H$ through a closed top quark loop. This contribution represents
$ g g \ra H g^* \ra H b \bar b$. We have not included this
contribution in the present paper as we do not consider it to be a
{\em genuine} $b \bar b H$ final state. This correction can be counted
as belonging to the {\em inclusive} $gg \ra H$ process. The same line
of reasoning has been argued in \cite{LH05-Higgs}. Nonetheless
from the experimental point of view it would be interesting to
include all these effects together with the NLO QCD corrections
and the electroweak corrections that we have studied here.

\vspace{2cm}
\noi {\bf Acknowledgments} \\
LDN expresses his gratitude and thanks to G. Altarelli and P.
Aurenche for their supervision, support, most helpful discussions
and comments. We benefited a lot from discussions with
J.P.~Guillet and P.~Slavich. We also acknowledge discussions with
G.~B\'elanger, DAO Thi Nhung, DO Hoang Son, J.~Ellis, J.~Fujimoto,
K.~Kato, Y.~Kurihara, M.~M\"uhlleitner, E.~Pilon, P.~Uwer and J.
Vermaseren. LDN acknowledges the receipt of a  {\em Rencontres du
Vietnam} scholarship. LDN is supported by the {\em Marie Curie
Early Stage Training Fellowship of the European Commission}.

\newpage

\renewcommand{\thesection}{\Alph{section}}
\setcounter{section}{0}
\renewcommand{\theequation}{\thesection.\arabic{equation}}
\setcounter{equation}{0}
\noi {\Large {\bf Appendices}}

\section{The helicity amplitude method }
\label{appendix-helicity}
\subsection{The method}
We use a combination of helicity amplitude methods as  described
in \cite{kleiss_stirling,maina} to calculate the total cross
section. In the following we only want to highlight some key
features that were most useful for our calculation, for details of
the method we refer to\cite{kleiss_stirling,maina}. For our
process $g(p_1,\la_1)+g(p_2,\la_2)\rightarrow
b(p_3,\la_3)+\bar{b}(p_4,\la_4)+H(p_5)$ where the particles are
denoted by their momentum $p_i$ and  helicity $\lambda_i$ we write
the corresponding helicity amplitude as $\ali$.

\def\slashepi{\epsilon_i\kern -.720em {/}}
\def\slashpi{p_i\kern -.670em {/}}

\bea \ali&=&\eps_{\mu}(p_1,\la_1;q_1)\eps_{\nu}(p_2,\la_2;q_2) {\cal
M}^{\mu \nu}(\la_3,\la_4), \crn
 {\cal M}^{\mu\nu}(\la_3,\la_4)&=&
\bar{u}(p_3,\la_3)\Gamma^{\mu\nu} v(p_4,\la_4). \label{amp_form}
\eea
$\Gamma^{\mu\nu}$ is a string of Dirac $\gamma$ matrices. These $\gamma$
matrices represent either interaction vertices or momenta from the
fermion propagators. In our case the interaction vertices are the
vectorial gluon vertices in which case they represent $\slashepi
\;$, the scalar Higgs vertex and at one-loop  the pseudo-scalar
Goldstone coupling. For the momenta, in our implementation we
re-express them in terms of the independent external momenta
$p_1,p_2,p_3,p_4$. This applies also to the loop momenta after the
reduction formalism of the tensor integrals has been performed.
The first step in the idea of the helicity formalism we follow is
to  turn  each of these $\gamma$ matrices (apart from the
pseudo-scalar and the trivial scalar) into a combination of spinor
function $u \bar u$. We therefore transform our helicity
amplitude into products of spinors such as the helicity amplitude
could be written like a product $\bar u\; u
 \bar u\; ...u \bar u \; v$ with the possible insertion of $\gamma_5$'s in the string.
The different $u$, $\bar u$, $v$ in the string we have written have
of course, in general, different arguments. Nonetheless one can
turn each spinor product of two adjacent $\bar u u$, etc into a
complex number written in terms of the momenta in our problem as
we will see.

\noi In the first step, for the momentum $\slashpi$ with
$p_i^2=m_i^2$ we use \beq \slashpi=u(p_i,-)\bar
u(p_i,-)+u(p_i,+)\bar u(p_i,+)-m_i. \eeq \noindent The
polarization vector of the initial gluon $i$,
$\eps_{\mu}(p_i,\la_i;q_i)$, is also first expressed in terms of
spinors such as
\bea\ep_\mu(p_i,\la_i;q_i)&=&\fr{\bar{u}(p_i,\la_i)\ga_\mu
u(q_i,\la_i)}{[4(p_i.q_i)]^{1/2}}, \label{eps12_def}
\eea
\noindent where $q_{i}$ is an  {\em arbitrary} reference vector
satisfying the following conditions
\bea q_i^2=0, \quad
p_i.q_i\neq 0\label{cond_refer_vector}
\eea
Gauge invariance (transversality condition) requires that the
cross sections are independent of the choice of the reference
vector as we will see later. This acts as an important check of
the calculation, see later.  It is not difficult to prove that the
choice (\ref{eps12_def}) satisfies all the conditions for a
transverse polarization vector. In particular,
\bea
p_i.\eps(p_i,\la_i)&=&0, \quad  \eps(p_i,\la_i).\eps(p_i,\la_i)=0,
\crn  \eps_{\mu}(p_i,-\la_i)&=&\eps_{\mu}(p_i,\la_i)^*, \quad
\eps(p_i,\la_i).\eps(p_i,-\la_i)=-1,
\eea
where the reference vector is not written down explicitly. $i=1,2$
and no sum over $i$ must be understood. Then for
$\slashepi=\epsilon_\mu \gamma^\mu$ one uses  the so-called
Chisholm identity
\bea\bar{u}(p,\la)\ga_\mu
u(q,\la)\ga^\mu=2[u(p,-\la)
\bar u(q,-\la)+u(q,\la)\bar u(p,\la)],  \label{chisholm}
\eea
where all the spinors in Eq.~\ref{chisholm} are for massless
states in view of the lightlike condition on the reference frame
vector and of course the momentum of the real gluon.

\noi With $U(p_i,\la_i)$ representing
either $u(p_i,\la_i)$ or $v(p_i,\la_i)$ one uses the general
formulae
\bea
\bar U(p_i,\la_i)U(p_j,\la_j)&=&\frac{A_{\la_i\la_j}(p_i,p_j)+M_i
B_{\la_i\la_j}(p_i,p_j)+M_j
C_{\la_i\la_j}(p_i,p_j)}{\sqrt{(p_i.k_0)(p_j.k_0)}},\crn \bar
U(p_i,\la_i)\gamma_5U(p_j,\la_j)&=&-\la_i
\frac{A_{\la_i\la_j}(p_i,p_j)-M_i B_{\la_i\la_j}(p_i,p_j)+M_j
C_{\la_i\la_j}(p_i,p_j)} {\sqrt{(p_i.k_0)(p_j.k_0)}}, \label{def_U5U}
\eea
where
\bea M_i&=&+m_i \,\,\,\,\, \text{if}\,\,\,\,\,
U(p_i,\la_i)=u(p_i,\la_i),\crn M_i&=&-m_i \,\,\,\,\,
\text{if}\,\,\,\,\, U(p_i,\la_i)=v(p_i,\la_i),\crn
A_{\la_i\la_j}&=&\delta_{\la_i -\la_j} \la_i
\left((k_0.p_i)(k_1.p_j)-(k_0.p_j)(k_1.p_i)-i \la_i
\epsilon_{\mu\nu\rho\sigma}k_0^\mu k_1^\nu p_i^\rho p_j^\sigma
\right), \crn B_{\la_i\la_j}&=&\delta_{\la_i\la_j}(k_0.p_j),\,\,
C_{\la_i\la_j}=\delta_{\la_i\la_j}(k_0.p_i),
\eea
with $k_{0,1}$ being auxiliary vectors such that $k_0^2=0$,
$k_1^2=-1$ and $k_0.k_1=0$. No sum over repeated indices must be
understood. For instance, we can choose $k_0=(1,0,1,0)$
and $k_1=(0,1,0,0)$. With this choice, it is obvious to see that
the denominator in (\ref{def_U5U}) can never vanish if the bottom mass is kept.
If one would like to neglect $m_b$, that choice can bring $p_3.k_0$ or
$p_4.k_0$ to zero in some cases. If this happens, one can tell the code to
choose $k_0=(1,0,-1,0)$ instead of the above choice. In fact, that is what we did
in our codes.

In the case of spinors representing a massless state, the helicity
formalism simplifies considerably. Only $A_{\la_i\la_j}$ is
needed. Traditionally we introduce the $C$-numbers $s(p,q)$ and
$t(p,q)$,
\bea s(p,q)\equiv\bar{u}(p,+)u(q,-)=A_{+-}(p,q),\,\,
t(p,q)\equiv\bar{u}(p,-)u(q,+)=-s(p,q)^*.
\eea
These are the functions that appear in our code for the massless
$b$ quark. The massless case is also used when expressing the
gluon polarisation vector to which we now turn.

\subsection{Transversality and gauge invariance}
The reference vector used for the polarisation of the gluon can be
changed at will. Changing the reference vector from $q$ to
$q^\prime$ amounts essentially to a gauge transformation. Indeed
one has \cite{kleiss_stirling}
\bea
\eps^\mu(p,\la;q^\prime)=e^{i\phi(q^\prime,q)}\eps^\mu(p,\la;q)+\beta(q^\prime,q)p^\mu,\,\,
\label{relation_eps}
\eea where
\bea
e^{i\phi(q^\prime,q)}&=&\left[\fr{s(p,q)}{t(p,q)}\fr{t(p,q^\prime)}{s(p,q^\prime)}\right]^{1/2},\crn
\beta(q^\prime,q)&=&\fr{2}{[4(q^\prime.p)]^{1/2}}\fr{t(q,q^\prime)}{t(q,p)}.
\eea
Therefore up to the phase factor, the difference is contained in
the momentum vector of the gluon. QCD gauge invariance for our
process leads to the important identity


\bea
|\aliq|^2=|\aliqp|^2,\label{check_gauge}\eea as long as
$q^\prime_{1,2}$ satisfy the condition (\ref{cond_refer_vector}).
We have carefully checked that the numerical result for the norm
of each helicity amplitude at various point in  phase space is
independent of the reference vectors $q_{1,2}$ up to 12 digits using double precision.
By default, our numerical evaluation is based on the
use of $q_{1,2}=(p_2,p_1)$. For the checks in the case of massive
$b$ quarks the result with $q_{1,2}=(p_2,p_1)$ is compared with
the one using any $q_{1,2}$ such as the conditions (\ref{cond_refer_vector}) are obeyed.
In the case of massless $b$ quarks it is simplest to take
$q_{1,2}=(p_3,p_4)$.

This check is a an important check on many ingredients that enter
the calculation: the Dirac spinors, the gluon polarization
vectors, the propagators, the Lorentz indices, the loop integrals.
It has been used extensively in our numerical calculation.

\renewcommand{\theequation}{\thesection.\arabic{equation}}
\setcounter{equation}{0}
%
\section{Optimisation}
\label{optimisation}

Each helicity amplitude ${\cal A}(\la_1,\la_2;\la_3\la_4) \equiv
\alio$, a C-number, is calculated numerically in the Fortran code.
The price to pay is that the number of helicity amplitudes to be
calculated can be large, $16$ in our case for the electroweak loop
part. Some optimisation is necessary. The categorisation of the
full set of diagrams into three  gauge invariant classes as shown
in section~\ref{section_3classes} is a first step. We have sought
to write each diagram as a compact product of blocks and
structures containing different properties of the amplitude. We
write the amplitude according to a colour ordering pattern that
defines three channels. The ordering is in a one-to-one
correspondance with the three channels or diagrams shown in
Fig.~\ref{diag_gg_LO}. The $T$-type is the direct channel, the
$U$-type is the  crossed one obtained from the $T$-type by
interchanging the two gluons and the $S$-type is the one involving
the triple gluon vertex. The helicity amplitude for each diagram
can thus be represented as
\bea
\alio^{T,U,S}=CME(a,b)\times Cc\times FFE\times SME(\la_i),
\eea
where
\begin{itemize}
\item
$CME(a,b)$ is the colour matrix element. $a,b$ are the colour
indices of the two initial gluons\footnote{Other colour indices of
the bottom quarks are omitted here for simplicity}. The colour
products  can be $(T^aT^b)$, $(T^bT^a)$ or $[T^a,T^b]$
corresponding to the $3$ $T$, $U$, $S$ channels respectively
\item
$Cc$ contains all the common coefficients like the strong coupling constant
$g_s$ or factors common to all diagrams and amplitudes such as the
normalisation factor entering the representation of the
polarisation vector of the gluon, see Eq.~\ref{eps12_def}
\item
$FFE$, form factor element, contains all the denominators of
propagators, loop functions as well as various scalar products of
external momenta $\{p_1,p_2,p_3,p_4\}$ {\it i.e.} all the scalar
objects which do not depend on the helicity $\la_i$
\item
$SME(\hat \la)$, standard matrix element, is a product of the
scalar spinor functions $A_{\la_i\la_j}$, $B_{\la_i\la_j}$ and
$C_{\la_i\la_j}$ defined in Appendix~\ref{appendix-helicity}.
\end{itemize}

For each channel, say $\alio^T$, the most complicated and
time-consuming part is the $FFE$. That is why we want to factorise
it out and put it in a common block so that in order to calculate
all the $16$ helicity configurations of $\alio^T$ we just need to
calculate $FFE$ once. This is done at every point in  phase space.
This kind of factorisation can be easily carried out in Form.\\
\noi $SME(\hat \la)$ is also complicated  because the bottom quark
is massive and $\gamma_5$ occurs in the ``helicity strings". Thus
we have to optimize this part as well. The way we do it for all
the $3$ groups is as follows. In Form, we have to find out all the
generic expressions of $SME(\hat \la)$. There are $12$ of them at
tree level and $68$ at one-loop if we choose $q_{1,2}=p_{2,1}$ for
the reference vectors. For instance,
\bea
SME_1&=&[\bar{u}(\la_3,p_3)v(\la_4,p_4)]\times [\eps_{\mu}(\la_1,p_1,p_2)p_4^{\mu}]\times [\eps_{\nu}(\la_2,p_2,p_1)p_4^{\nu}],\crn
&=&BME_1(\la_3,\la_4)\times BME_2(\la_1)\times BME_3(\la_2),
\eea
can be expressed in terms of  $3$ basic matrix elements ($BME$).
Each $BME$ occurs several times when calculating all the $SME(\hat
\la)$. The number of $BME$ is $31$. Each $BME$ is written in terms
of scalar spinor functions $A_{\la_i\la_j}$, $B_{\la_i\la_j}$,
$C_{\la_i\la_j}$. All the $SME$ or $BME$ can be found and
abbreviated in Form. As an alternative, we can use Perl for such
an operation. The Form output is converted directly into a Fortran
code
for numerical evaluation. Needless to say, all the abbreviations of $SME$ or $BME$ must be put in common blocks.\\

To get the final result, we have to sum over all the channels. The
grouping can be re-arranged in terms of an Abelian part and a
non-Abelian part according to
\bea \alio&=&
\alio^{T}+\alio^{U}+\alio^{S}\,,\crn
&\equiv&\{T^a,T^b\}\alio^{Abel}+[T^a,T^b]\alio^{NAbel}\,,\eea
where \bea \alio^{Abel}&=&\fr{1}{2}(\alio^T+\alio^U)\,,\crn
\alio^{NAbel}&=&\alio^S+\fr{1}{2}(\alio^T-\alio^U)\,\eea
corresponding to the Abelian and Non-Abelian parts respectively.
The amplitude squared then contains no interference term between
the Abelian and Non-Abelian parts: \bea\mid
\alio\mid^2=\fr{1}{256}\left(\fr{28}{3}\mid
\alio^{Abel}\mid^2+12\mid \alio^{NAbel}\mid^2\right)\,\eea where
$\fr{1}{256}=\fr{1}{4}\times\fr{1}{8}\times\fr{1}{8}$ is the spin-
and colour- averaging factor.

\newpage


\end{document}